\documentstyle[aps,twocolumn]{revtex}
\begin{document}
\title{Quantum State Discrimination}
\author{Anthony Chefles}
\address{Department of Physical Sciences, \\ University of Hertfordshire,
       Hatfield AL10 9AB, Herts, UK \\ email: A.Chefles@herts.ac.uk}
\input epsf
\epsfverbosetrue
\maketitle
\begin{abstract}
There are fundamental limits to the accuracy with which one can
determine the state of a quantum system.  I give an overview of
the main approaches to quantum state discrimination. Several
strategies  exist.  In quantum hypothesis testing, a quantum
system is prepared  in a member of a known, finite set of states,
and the aim is to guess which one with the minimum probability of
error. Error free discrimination is also sometimes possible, if we
allow for the possibility of obtaining inconclusive results. If no
prior information about the state is provided, then it is
impractical to try to determine it exactly, and it must be
estimated instead.  In addition to reviewing these various
strategies, I describe connections between state discrimination,
the manipulation of quantum entanglement, and quantum cloning.
Recent experimental work is also discussed.
\end{abstract}
\section{Introduction}
\renewcommand{\theequation}{1.\arabic{equation}}
\setcounter{equation}{0} The information which describes the state
of a physical system is that which is required to predict its
future evolution and its effect on other systems. Among the other
physical systems whose dynamics it can influence are measuring
devices. The state then contains all information that can be
extracted by measurement and thus, all information that we can
acquire about the system.

The state of a classical system is described by dynamical variables.  For a one
dimensional point particle, these are its position {\em q} and momentum {\em p}. If we
have complete knowledge of the values of these variables, and also of the equations that
describe their temporal evolution, then we can predict, with arbitrarily high accuracy,
the future state of the system, and how it will interact with other systems.

The assumption that we can measure these variables with
arbitrarily high accuracy is, however, an idealisation.  In
addition to the fact that the results of any real experiment will
be affected by some uncontrollable noise, our measuring devices
have finite precision and can only record a finite amount of
information. This implies that they will not be able to record
exactly the values of continuous variables which form the basis of
the descriptions of most classical systems. Often then, we must
settle for an approximate description of the state. In our simple
example of a one dimensional point particle, practical limitations
on our ability to measure {\em q} and {\em p} precisely might lead
us to use a joint probability density ${\varrho}(q,p)$ instead.
Often, such a description is adequate. However, there do exist
situations where even the smallest, finite uncertainty in our
knowledge of the state variables will be amplified over time to
such large proportions that long term prediction of the behaviour
of the system becomes impossible. As is well-known, this extreme
sensitivity to initial conditions is one of the chief hallmarks of
the phenomenon of dynamical chaos\cite{Chaos}.

As a matter of principle though, there are no {\em fundamental}
limitations on the precision with which we can determine the state
variables of a classical system. This is because the state
variables are also {\em observables}.  The amount of information
that we can acquire about observables is always increasing as more
refined measurements become possible. This equivalence of state
information and the information that is potentially accessible
through measurement is a highly non-trivial feature of classical
physics which, despite its transparent nature, should not be taken
for granted. Classical physics is only an approximate description
of our world, and currently, our most accurate description of it
is provided by quantum theory.  Here, states and observables are
completely distinct concepts.

The state of a quantum system is represented by a normalised
vector $|{\psi}{\rangle}$ in a complex, linear vector space. The
observable properties of quantum systems, by contrast, are the
same as those of classical systems: position, momentum etc. Unlike
their classical counterparts, they are not represented by simple
numerical variables which evolve deterministically over time.
Instead, they are represented by Hermitian operators on the vector
space. We shall make no distinction in what follows between
observables and their corresponding operators, and shall denote
both aspects of a generic observable by ${\Omega}$.

Knowledge of the state vector is instrumental in making
predictions about the outcome of measuring an observable
${\Omega}$. As is discussed in every introductory quantum
mechanics text, e.g. \cite{Bransden}, these predictions are not
generally of a deterministic nature, and are concerned instead
with statistical quantities.  The average, or expectation value of
${\Omega}$ for an ensemble of quantum systems all prepared with
the same state vector $|{\psi}{\rangle}$ is given by the inner
product of $|{\psi}{\rangle}$ and ${\Omega}|{\psi}{\rangle}$,
\begin{equation}
{\langle}{\Omega}{\rangle}={\langle}{\psi}|{\Omega}|{\psi}{\rangle}.
\end{equation}
In order to treat a single system, we express ${\Omega}$ in terms
of its eigenvalues ${\omega}_{k}$ and orthogonal eigenstates
$|{\omega}_{k}{\rangle}$:
\begin{equation}
{\Omega}=\sum_{k}{\omega}_{k}|{\omega}_{k}{\rangle}{\langle}{\omega}_{k}|.
\end{equation}
The eigenvalues ${\omega}_{k}$ are the values that the observable
${\Omega}$ can take. Inserting Eq. (1.2) into Eq. (1.1), we obtain
\begin{equation}
{\langle}{\Omega}{\rangle}=\sum_{k}{\omega}_{k}|{\langle}{\psi}|{\omega}_{k}{\rangle}|^{2}.
\end{equation}
The expectation value of any quantity is equal to the sum of the
values that it can assume, multiplied by their respective
probabilities.  From this, we infer that the probability
$P({\omega}_{k}|{\psi})$ of obtaining the result ${\omega}_{k}$
for the state $|{\psi}{\rangle}$ is
\begin{equation}
P({\omega}_{k}|{\psi})=|{\langle}{\psi}|{\omega}_{k}{\rangle}|^{2}.
\end{equation}
If the result ${\omega}_{k}$ is obtained, then the state of the
system following the measurement will become
$|{\omega}_{k}{\rangle}$.  This is the contentious process of
state vector reduction, or `collapse of the wavefunction'.  Such a
measurement is known as a von Neumann measurement.

The vector $|{\psi}{\rangle}$ is the most complete specification
of the state of a quantum system.  As is the case in classical
physics, it is not always possible to specify the state exactly.
By analogy with the probability density ${\varrho}(q,p)$, in
dealing with quantum systems we will sometimes have to use a
probability distribution for the state vector itself.  This is
described by a {\em density operator},
\begin{equation}
{\rho}=\sum_{r}p_{r}|{\psi}_{r}{\rangle}{\langle}{\psi}_{r}|.
\end{equation}
Here, $p_{r}$ is the probability that the state vector of the
system is $|{\psi}_{r}{\rangle}$, and $\sum_{r}p_{r}=1$.  When a
quantum system is known to have a specific state vector, it is
said to be in a {\em pure state}. Otherwise, it is in a {\em mixed
state}. The generalisation of Eq. (1.4) for the probability of
obtaining result ${\omega}_{k}$ when the system is in a possibly
mixed state ${\rho}$ is
\begin{equation}
P({\omega}_{k}|{\rho})={\langle}{\omega}_{k}|{\rho}|{\omega}_{k}{\rangle}={\mathrm
Tr}{\rho}|{\omega}_{k}{\rangle}{\langle}{\omega}_{k}|.
\end{equation}
Here, `Tr' stands for the trace operation.  The trace of an
operator is, in a matrix representation, the sum of its diagonal
elements.  The numerical value of this sum is basis independent.
So, if we choose the basis to be that formed by the eigenstates
$|{\omega}_{k}{\rangle}$ of ${\Omega}$, then the trace is seen to
be the sum of the corresponding eigenvalues ${\omega}_{k}$. One
can quite easily show, using Eq. (1.5), that ${\mathrm
Tr}{\rho}=1$. From Eq. (1.6), it follows that the expectation
value of ${\Omega}$ is
\begin{equation}
{\langle}{\Omega}{\rangle}=\sum_{k}{\omega}_{k}{\langle}{\omega}_{k}|{\rho}|{\omega}_{k}{\rangle}={\mathrm
Tr}{\rho}{\Omega}.
\end{equation}
 The evident distinction between states
and observables in quantum mechanics begs the question: to what extent can we determine
the state of a quantum system?  The ability to do this would confer many benefits.  The
most obvious of these is that we would be able to evaluate the probability distribution
for the results of any measurement that we might wish to perform upon a quantum system
about whose state we have no prior information. Another less obvious benefit, which we
shall later see, is that it would allow two parties to communicate across arbitrarily
large distances, instantaneously, in violation of the special theory of relativity.

The state itself is not an observable in quantum mechanics.  As it
happens, the impossibility of measuring the state has benefits of
its own. As was demonstrated initially by Bennett and
Brassard\cite{BB}, the impossibility of measuring
$|{\psi}{\rangle}$ precisely permits the existence of provably
secure protocols for the transmission of confidential information.
The security of quantum cryptographic protocols, unlike that of
classical ones, is a consequence of physical
theory\cite{Phoenix,Hughes}. The security of classical protocols
depends upon unproven assumptions about the complexity of the
decoding problem.  One of the most widely used cryptosystems, the
RSA cryptosystem, exploits the difficulty of the problem of
reducing a large number to its set of prime factors \cite{RSA}. No
efficient classical algorithm for carrying out this task has been
discovered.  However, it has not been proven that one does not
exist.  In fact, it has recently been shown by Shor that an
efficient algorithm does exist for quantum
computers\cite{Shor,Barenco}.

In this article, we will examine the problem of determining the
state of a quantum system. Although the state is not, strictly
speaking, an observable, through a judicious choice of legitimate
observables, we can obtain information about it. Several
strategies for state discrimination exist.  The one we would use
in any particular situation depends upon the type of information
about the state we wish to obtain, and also on any prior
information that we might possess.

The first strategy we examine is {\em quantum hypothesis testing}.
Here, we are given a system whose state belongs to a known, finite
set. Our aim is to guess, with the minimum probability of error,
which of these states the system is in.

In the course of our discussion, we shall encounter the elegant
formalism of generalised quantum measurements.  This is based on a
few necessary and sufficient conditions which any physically
possible operation on a quantum system must satisfy.

Sometimes errors can be avoided altogether if we allow for the
possibility of inconclusive results.  We will see how this can be
achieved. Another scenario is when the state does not belong to
some finite, known set, and can be any state in the entire vector
space.   Under these circumstances, the set of possible states is
infinite.  Since any measurement can record only a finite amount
of information, it is necessary to consider only a finite, but
suitably large set of possible states.  These states are known as
{\em guess} states.  An appropriate measurement strategy is one
which selects the guess state which the actual state most closely
resembles, as often as possible. This is known as {\em quantum
state estimation}.

We shall also explore the relationship between trying to
discriminate between quantum states and other matters, such as the
problem of cloning quantum states, and the manipulation of
nonlocal correlations between quantum systems, which can exist due
to the strange phenomenon of quantum entanglement.
\section{Quantum Hypothesis Testing}
\renewcommand{\theequation}{2.\arabic{equation}}
\subsection{Basic strategy}
\setcounter{equation}{0} In this article, we shall see that
several approaches to the problem of quantum state discrimination
exist.  All of them refer to the same basic scenario.  One party,
conventionally called Alice, prepares a quantum system in a member
of a set of quantum states.  She might not prepare all of these
states with the same probability. She then passes the system onto
her colleague, Bob.  His task is obtain as much information about
the state which she prepared as he possibly can.  Here, we use the
term `information' in a broad sense, and do not refer as yet to
any specific measure. The main differences between the state
discrimination strategies considered in this article correspond to
the different types of information that Bob might wish to obtain.

In this section, we will assume that some information about the state is given to Bob.  He
is told what the possible states of the system are, and he is also told the probability
that the system was prepared in each of them.  Here, we will consider only the situation
where there are $N$ possible states, represented by density operators ${\rho}_{j}$, where
$j=1,{\ldots},N$, for some finite $N$.  Bob is also told the probability, ${\eta}_{j}$,
that the system was prepared in each of them.  These probabilities, known as the {\em a
priori} probabilities, satisfy
\begin{equation}
\sum_{j=1}^{N}{\eta}_{j}=1,
\end{equation}
since the system will, with certainty, be prepared in one of the
states ${\rho}_{j}$.

Historically, the first strategy for state discrimination was that
advanced by Helstrom\cite{Helstrom}.  This strategy is known as
{\em quantum hypothesis testing}.  In his attempt to determine the
state of the system, Bob performs some measurement. The key
feature of quantum hypothesis testing, as opposed to some other
strategies, in particular, the one we shall examine in the next
section, is that after his measurement, he is required, on the
basis of his experimental results, to make a decision as to what
the state was. He is not allowed to say `don't know'.  We will see
that, if the states are not orthogonal, then no test exists which
allows him to guess correctly all of the time, so that there will,
in general, be a non-zero probability of error, which we shall
denote by $P_{E}$.  Likewise, we will write the probability of
correctly determining the state as $P_{D}=1-P_{E}$.

Since there are $N$ states, his experiment must have $N$ outcomes,
which we call ${\omega}_{k}$.  Following this kind of test, if Bob
obtains the result ${\omega}_{j}$, he makes the hypothesis that
the state given to him by Alice was ${\rho}_{j}$.

To determine the probability of error, Bob needs to know the a
priori probability ${\eta}_{j}$ of being given the state
${\rho}_{j}$  and the probability, given that ${\rho}_{j}$ was
sent, that he will obtain the result ${\omega}_{k}$, for all
$j,k$. The probabilities form the {\em channel matrix}
$[P({\omega}_{k}|{\rho}_{j})]$.  The elements of this matrix
satisfy the {\em completeness} condition
\begin{equation}
\sum_{k=1}^{N}P({\omega}_{k}|{\rho}_{j})=1.
\end{equation}
This expresses the fact that, no matter which ${\rho}_{j}$ Bob
receives, his measurement will, with certainty, yield one of the
outcomes ${\omega}_{k}$. The total error probability $P_{E}$ is
found to be
\begin{equation}
P_{E}=1-P_{D}=1-\sum_{j=1}^{N}{\eta}_{j}P({\omega}_{j}|{\rho}_{j}).
\end{equation}

Quantum hypothesis testing actually belongs to a more general
class of strategies known as quantum Bayes'
strategies\cite{Helstrom}. The general quantum Bayes strategy
assigns a cost $C_{kj}$ to making hypothesis ${\omega}_{k}$ when
the state was ${\rho}_{j}$.  The coefficients $C_{kj}$  are known
as the elements of the {\em Bayes' cost matrix}. The scenario can
easily be understood in terms of gambling. Alice sends Bob one of
the states ${\rho}_{j}$. If Bob says `${\omega}_{k}$', then he
must pay Alice $C_{kj}$ currency units. Some elements of the cost
matrix can be negative, in which case Alice pays Bob, enabling him
to win money. The average amount that Bob will pay Alice is then
given by the Bayes' cost function
\begin{equation}
C_{B}=\sum_{jk}{\eta}_{j}C_{kj}P({\omega}_{k}|{\rho}_{j}).
\end{equation}
For a fixed cost matrix $[C_{kj}]$ and a priori probabilities
${\eta}_{j}$, Bob's task is to minimise the overall Bayes' cost
$C_{B}$, that is, to use a measurement which minimises his average
payout to Alice.  The only quantities that Bob is free to vary are
the channel matrix elements $P({\omega}_{k}|{\rho}_{j})$. Since
the possible states ${\rho}_{j}$ are fixed, the only thing that
Bob is free to vary is his measurement strategy.

The form of the cost matrix depends on the particular situation.  In general, some errors
may be more costly than others.  If every error has equal cost, then the corresponding
Bayes' cost is closely related to the average error probability.  There is no cost when
the result is correct, so the diagonal elements of the cost matrix, $C_{jj}$, are zero.
Let all other elements have an associated cost {\em c}: that is, $C_{jk}=c$ for
$j{\neq}k$.  Then one can show, using the definitions of the error probability and Bayes'
cost in Eqs. (2.3) and (2.4), and the completeness condition in  Eq. (2.2), that the
Bayes' cost and error probability are related by $C_{B}=cP_{E}$.  When all errors have the
same cost $c$, minimisation of the Bayes' cost is equivalent to minimisation of the error
probability.

The lowest value of $P_{E}$ is obtained by varying the elements of
the channel matrix, $P({\omega}_{k}|{\rho}_{j})$.   As a
consequence of the non-trivial nature of the measurement process
in quantum mechanics, the form of this matrix cannot be specified
arbitrarily.  Consider for example a von Neumann measurement of an
observable ${\Omega}$. Let us assume that the states ${\rho}_{j}$
are pure states $|{\psi}_{j}{\rangle}{\langle}{\psi}_{j}|$.  We
would like to associate each outcome with a unique state, so that
if Bob obtains the result ${\omega}_{j}$ he will make the
hypothesis that the state Alice sent him was
$|{\psi}_{j}{\rangle}$.  Clearly, the number $N$ of possible
states must be equal to the number of outcomes. Consequently, we
must also then assume that the $|{\psi}_{j}{\rangle}$ are linearly
independent.  Otherwise, there would be more states than outcomes.

In the Introduction, we saw that each outcome of a von Neumann
measurement corresponds to a different eigenvalue of an Hermitian
operator ${\Omega}$.  The eigenvalues ${\omega}_{j}$ are the
possible numerical values of ${\Omega}$, considered as an
observable property of the system.  From Eq. (1.4), we see that
the channel matrix elements are given by the square-overlaps
between the $|{\psi}_{j}{\rangle}$ and the eigenstates of
${\Omega}$:
\begin{equation}
P({\omega}_{k}|{\psi}_{j})=|{\langle}{\psi}_{j}|{\omega}_{k}{\rangle}|^{2}.
\end{equation}
The diagonal elements of this matrix must all be equal to 1 if the
error probability is to vanish.  This clearly gives the
requirement that $|{\psi}_{j}{\rangle}=|{\omega}_{j}{\rangle}$,
which cannot be the case if the $|{\psi}_{j}{\rangle}$ are
non-orthogonal.

A simple von Neumann measurement of this kind, however, is often
the most useful for the kind of strategy we are considering.  In
fact, it was proven in 1973 by Kennedy\cite{Kennedy} that if we
are attempting to distinguish between $N$ pure states which are
linearly independent, as we have been assuming, then there is
always a von Neumann measurement which is optimal, in the sense
that it can be used to obtain the smallest possible value of
$P_{E}$.  It follows that only orthogonal states can be perfectly
discriminated.
\subsection{Hypothesis testing for two pure states}

The simplest set of linearly independent pure states, and
historically the first set for which an explicit expression for
the minimum error probability was obtained, is that of just two
states.  The problem of finding the minimum value of $P_{E}$ for
two pure states, which we shall simply simply by
$|{\psi}_{\pm}{\rangle}$, was solved by Helstrom\cite{Helstrom}
and can be considered to be a pioneering work in quantum detection
theory.  Helstrom's optimal value of $P_{E}$ is
\begin{equation}
P_{E}({\mathrm
opt})=\frac{1}{2}\left(1-{\sqrt{1-4{\eta}_{+}{\eta}_{-}|{\langle}{\psi}_{+}|{\psi}_{-}{\rangle}|^{2}}}\right).
\end{equation}
Naturally, we would like to determine the von Neumann measurement
which can be used to attain this limit.  The corresponding basis
 states $|{\omega}_{\pm}{\rangle}$ have quite a simple form if we take the
 states $|{\psi}_{\pm}{\rangle}$ to be
\begin{equation}
|{\psi}_{\pm}{\rangle}={\cos}{\theta}|+{\rangle}{\pm}{\sin}{\theta}|-{\rangle}
\end{equation}
for some angle $0{\le}{\theta}{\le}{\pi}/4$, and where
$|{\pm}{\rangle}$ is an orthogonal basis for the space spanned by
$|{\psi}_{\pm}{\rangle}$.  For the states $|{\psi}_{\pm}{\rangle}$
in Eq. (2.7), the optimum detector states
$|{\omega}_{\pm}{\rangle}$ are
\begin{equation}
|{\omega}_{\pm}{\rangle}=\frac{1}{\sqrt
2}\left[\sqrt{1{\pm}{\xi}}|+{\rangle}{\pm}\sqrt{1{\mp}{\xi}}|-{\rangle}\right].
\end{equation}
Here,
${\xi}={\Delta}{\cos}2{\theta}/\sqrt{1+{\cos}^{2}2{\theta}({\Delta}^{2}-1)}$
where ${\Delta}={\eta}_{+}-{\eta}_{-}$. For alternative
expressions for the optimal detector states, see \cite{Tamagawa}.
The optimum detection strategy for a pair of {\em mixed} quantum
states has also been obtained.  For a full discussion,
see\cite{Osaki}.

 For two pure states with equal a priori
probabilities ${\eta}_{+}={\eta}_{-}=1/2$, the optimum detector
states in Eq. (2.8) have a much simpler form. and the optimum
measurement has appealing geometrical properties. When the a
priori probabilities are equal, we have ${\Delta}=0$, which in
turn implies that ${\xi}=0$.  The states
$|{\omega}_{\pm}{\rangle}$ are then given simply by

\begin{equation}
|{\omega}_{\pm}{\rangle}=\frac{|+{\rangle}{\pm}|-{\rangle}}{\sqrt{2}}
\end{equation}
and the minimum error probability simplifies to $P_{E}({\mathrm
opt})=[1-(1-|{\langle}{\psi}_{+}|{\psi}_{-}{\rangle}|^{2})^{1/2}]/2$.
All 4 states in Eqs. (2.7) and (2.9) are depicted in Figure 1. The
symmetrical properties of the measurement states
$|{\omega}_{\pm}{\rangle}$ with respect to the possible states
$|{\psi}_{\pm}{\rangle}$ are clearly visible in the figure.  The
$|{\omega}_{\pm}{\rangle}$ are as close as they can be to the
$|{\psi}_{\pm}{\rangle}$ whilst maintaining orthogonality.  The
reflection symmetry about the $|+{\rangle}$-axis is due to the
equality of the a priori probabilities ${\eta}_{\pm}$.  We can
also see from the figure that errors are unavoidable, since
$|{\omega}_{\pm}{\rangle}$ is not orthogonal to
$|{\psi}_{\mp}{\rangle}$.

The Helstrom measurement has recently been carried out in the
laboratory by Barnett and Riis\cite{BRiis}.  In this experiment,
the two states were non-orthogonal photon polarisation states,
having the form shown in Eq. (2.7), where the orthogonal states
$|+{\rangle}$ and $|-{\rangle}$ were the horizontal
$|{\leftrightarrow}{\rangle}$ and vertical
$|{\updownarrow}{\rangle}$ polarisation states respectively. The
experimental arrangement used is shown in Figure 2.  Pulses of
light emerged from the left in the horizontally polarised state
$|{\leftrightarrow}{\rangle}$.  These were then heavily attenuated
to the point where, on average, only 1 in 10 pulses contains a
photon. This was done to make the probability of there being 2 or
more photons per pulse negligible.  A Glan-Thompson polariser GTP
was then used to transform the photons into one of the states
$|{\psi}_{\pm}{\rangle}$ in Eq. (2.7).  The beam was then analysed
at a polarising beam splitter PBS oriented at an angle of
${\pi}/4$ to the horizontal.  To understand the action of this
beamsplitter, we refer to the states $|{\omega}_{\pm}{\rangle}$ in
Eq. (2.9), and again make the identifications
$|+{\rangle}=|{\leftrightarrow}{\rangle}$ and
$|-{\rangle}=|{\updownarrow}{\rangle}$.   A photon in the state
$|{\omega}_{+}{\rangle}$ would be transmitted by the beam
splitter, while $|{\omega}_{-}{\rangle}$ would be reflected.  The
transmitted and reflected states were fed to photodetectors
$D_{+}$ and $D_{-}$ respectively. Correct results were obtained
when a photon prepared in the state $|{\psi}_{j}{\rangle}$ was
detected at $D_{j}$, where $j={\pm}$.  If the photon was detected
at the other `wrong' detector, an error ensued.

In the Barnett-Riis experiment, both states had equal a priori
probabilities.  The minimum error probability is then
\begin{equation}
P_{E}({\mathrm opt})=\frac{1}{2}(1-{\sin}2{\theta}).
\end{equation}
Experimental results for the error probability for various values
of ${\theta}$ are shown alongside the theoretical minimum in
Figure 3.

Although the problem of finding the minimum error probability
$P_{E}({\mathrm opt})$ for two states has been solved completely,
it is generally difficult to find analytic expressions for more
than two states. Standard von Neumann measurements of the kind
which are optimal for two pure states, or indeed, any number of
linearly independent states, cannot detect all of the states if
they form a linearly dependent set. If the state space of the
system is $N$ dimensional, then a von Neumann measurement can have
at most $N$ outcomes. If there are more than $N$ states, then some
of these cannot be detected by the measurement.

Fortunately, the formalism of quantum mechanics does not restrict
us to state transformations described by von Neumann measurements.
This much is obvious from the transformation generated by the free
evolution of a quantum system.  This is described by the
Schr\"odinger equation, and is quite unlike what happens during a
measurement. In order to decide whether or not a given operation
on the quantum state is physically realisable, it would be helpful
to know what the general criteria are. These are firmly
established, and form the basis for the elegant formalism of {\em
generalised quantum measurements}\cite{Kraus}, which we now
describe.
\subsection{Generalised measurements}

Consider a quantum system initially prepared in the state
${\rho}$. An operation ${\rho}{\rightarrow}L({\rho})$ is carried
out of the system. This operation has $K$ distinguishable outcomes
which, as before, we label ${\omega}_{k},\;k=1,{\ldots},K$, with
corresponding final density operators ${\rho}'_{k}$.  In a von
Neumann measurement, the probability of outcome ${\omega}_{k}$ is
given by Eq. (1.6), that is, the trace of the product of the
initial density operator and
$|{\omega}_{k}{\rangle}{\langle}{\omega}_{k}|$.  In the more
general kind of measurement we describe here, the latter operators
are replaced by more general operators, known as {\em quantum
detection operators}, ${\Pi}_{k}$.  By analogy with Eq. (1.6), the
probability of obtaining result ${\omega}_{k}$ given the initial
state ${\rho}$, is
\begin{equation}
P({\omega}_{k}|{\rho})={\mathrm Tr}{\rho}{\Pi}_{k}.
\end{equation}
If ${\rho}$ is a pure state $|{\psi}{\rangle}{\langle}{\psi}|$,
then this probability $P({\omega}_{k}|{\psi})$ is simply
\begin{equation}
P({\omega}_{k}|{\psi})={\langle}{\psi}|{\Pi}_{k}|{\psi}{\rangle}.
\end{equation}
Naturally, $P({\omega}_{k}|{\psi})$ is always real.  This implies
that the quantum detection operators must be Hermitian.  This
probability must also be non-negative for all states.  Thus, the
expectation value of ${\Pi}_{k}$ must always be non-negative.
Operators whose expectation values are non-negative for all
possible states are said to be positive (semi-definite).  They may
be equivalently defined as operators whose eigenvalues are
non-negative.  One further constraint on the form of these
operators comes from the requirement that the possible outcomes
${\omega}_{k}$ are exhaustive, which implies that
$\sum_{k}P({\omega}_{k}|{\rho})=1$ for all possible states.  From
this, it follows that the ${\Pi}_{k}$ form a {\em resolution of
the identity},
\begin{equation}
\sum_{k}{\Pi}_{k}=1.
\end{equation}
The conditions we have just given are the necessary and sufficient
conditions for the realisability of an experiment whose outcomes
have the probability distribution
$P({\omega}_{k}|{\rho})$\cite{Kraus}. Such an operation is also
commonly known as a {\em positive operator-valued measure}
operation, or POVM, and the detection operators are called the
elements of the POVM.

Often, and particularly in state discrimination, we are only
interested in these probabilities, and not overly concerned about
how the state of the system is transformed by the measurement.
However, this is not always the case, indeed we shall be concerned
about this issue in section IV.  It is then useful to know what
form this state transformation must take.  To this end, consider
the operator
\begin{equation}
A_{k}=U_{k}{\Pi}_{k}^{1/2},
\end{equation}
where $U_{k}$ is any unitary operator.  From this expression, and
from the fact that $U_{k}^{\dagger}U_{k}=1$, we can see that
${\Pi}_{k}=A_{k}^{\dagger}A_{k}$, and that the detection
probability $P({\omega}_{k}|{\rho})$ can be alternatively
expressed as ${\mathrm Tr}A_{k}^{\dagger}A_{k}{\rho}$.  It can
also be expressed as ${\mathrm Tr}A_{k}{\rho}A_{k}^{\dagger}$,
since the trace of a product of operators is invariant under
cyclic permutations.  The post-measurement density operator, given
that result ${\omega}_{k}$ was obtained, is
\begin{equation}
{\rho}'_{k}=\frac{A_{k}{\rho}A^{\dagger}_{k}}{P({\omega}_{k}|{\rho})}.
\end{equation}
The presence of the probability in the denominator serves to give
${\mathrm Tr}{\rho}'_{k}=1$, normalising the state.  If we do not
actually record the result of the measurement, then the final
density operator, which may simply be denoted by ${\rho}'$, is
given by a distribution of the density operators ${\rho}_{k}^{'}$
corresponding to the possible outcomes of the operation, weighted
by their respective probabilities $P({\omega}_{k}|{\rho})$.  That
is
\begin{equation}
{\rho}'=\sum_{k}P({\omega}_{k}|{\rho}){\rho}'_{k}=\sum_{k}A_{k}{\rho}A^{\dagger}_{k}.
\end{equation}
The formalism we have outlined appears to be more general than the
description of quantum state changes given in introductory quantum
mechanics texts. There, usually only unitary operations and von
Neumann measurements are discussed.  It is easy to see that these
operations are special cases of the generalised measurements we
have just described.  A von Neumann measurement of an operator
${\Omega}$ with the orthogonal eigenstates
$|{\omega}_{k}{\rangle}$ can be expressed in terms of the
operators $P_{k}=|{\omega}_{k}{\rangle}{\langle}{\omega}_{k}|$.
These are projection operators, and are clearly Hermitian and
positive, having eigenvalues 0 and 1. They also form a resolution
of the identity
\begin{equation}
\sum_{k}P_{k}=1,
\end{equation}
which expresses the completeness of the orthogonal basis
$|{\omega}_{k}{\rangle}$.  They are also idempotent, that is,
$P_{k}^{2}=P_{k}$.  It follows that the projectors $P_{k}$ satisfy
the properties required for them to be physical transformation
operators $A_{k}$. Applying Eq. (2.15), we see that the possible
post-measurement states are just the eigenstates
$|{\omega}_{k}{\rangle}$, in accordance with the idea of collapse
of the wavefunction.  If the measurement result is not recorded,
substituting the $P_{k}$ for $A_{k}$ in Eq. (2.16) just gives a
statistical mixture of the eigenstates, weighted by their
respective probabilities $P({\omega}_{k}|{\rho})$.

At the other extreme, if only one $A_{k}$, which we may just call
$A$, is non-zero, then the resolution of the identity in Eq.
(2.13) implies that $A^{\dagger}A=1$, i.e. that $A$ is unitary.
Equations (2.13) and (2.16) are both equivalent here, since there
is only one `outcome', which represents unitary evolution
according to the von Neumann equation
\begin{equation}
i{\hbar}\frac{d{\rho}}{dt}=[H,{\rho}],
\end{equation}
where $[H,{\rho}]$ is the commutator $H{\rho}-{\rho}H$.  This is
simply the generalisation of the Schr\"odinger equation to cover
mixed states.  The solution is ${\rho}(t)=U{\rho}(0)U^{\dagger}$,
where $U={\mathrm e}^{\frac{-iHt}{\hbar}}$.  Any unitary operator
$U$ can be written in this form for some Hamiltonian {\em H}.  So,
if we are sufficiently able to tailor the Hamiltonian $H$ of our
system, then we can generate any unitary evolution.

Despite its appearance, the generalised measurement formalism is
not really more general than these two more familiar types of
state transformation.  An important result, known as the Naimark
theorem\cite{Naimark}, tells us that {\em any} generalised
measurement can be realised with an ancillary system, a unitary
operation and a von Neumann measurement. Specifically, if we wish
to realise a generalised measurement with $K$ outcomes, we need a
large ancillary system. The system of interest is then made to
interact unitarily with the ancilla.  In general, this results in
the original system and the ancilla becoming {\em entangled}.
Entanglement is a feature of quantum mechanics we shall examine at
greater length in section IV.  Following this interaction, a von
Neumann measurement is performed on the ancilla.  As a consequence
of this entanglement, this measurement also transforms the state
of our original system, and the results of this measurement give
rise to the corresponding transformations in Eq. (2.15).  The
effect of a generalised measurement, implemented by a unitary
interaction with an ancilla, followed by a measurement on the
latter, is illustrated in Figure 5.
\subsection{Hypothesis testing for multiple states}

Returning now to the problem of state discrimination, Bob has in
his possession a quantum system prepared in one of the $N$ states
${\rho}_{j}$, with a priori probabilities ${\eta}_{j}$.  His aim
is to determine the strategy with $N$ outcomes whose detection
operators ${\Pi}_{k}$ give the minimum value of the error
probability.  If outcome ${\omega}_{j}$ is taken to correspond to
detection of the state ${\rho}_{j}$, then the minimum value of the
error probability is obtained from the fact that the probability
of correctly identifying the state ${\rho}_{j}$ will be ${\mathrm
Tr}{\rho}_{j}{\Pi}_{j}$. The sum of these probabilities for the
{\em N} states ${\rho}_{j}$, weighted by their a priori
probabilities ${\eta}_{j}$, gives the total probability $P_{D}$
that the state will be correctly identified. The error probability
$P_{E}$ is equal to $1-P_{D}$, giving
\begin{equation}
P_{E}=1-\sum_{j}{\eta}_{j}{\mathrm Tr}{\Pi}_{j}{\rho}_{j}.
\end{equation}
Holevo\cite{Holevo} and Yuen {\em et al}\cite{Yuen} independently
determined the necessary and sufficient conditions that a set of
detection operators must satisfy to give the minimum value of
$P_{E}$. These are
\begin{eqnarray}
{\Pi}_{j}[{\eta}_{j}{\rho}_{j}-{\eta}_{k}{\rho}_{k}]{\Pi}_{k}&=&0,
\\ * {\Gamma}-{\eta}_{j}{\rho}_{j}&{\ge}&0,
\end{eqnarray}
where we have defined an operator ${\Gamma}$ known as the {\em
Lagrange operator}
\begin{equation}
{\Gamma}=\sum_{k}{\eta}_{k}{\Pi}_{k}{\rho}_{k},
\end{equation}
which, as a consequence of the condition in Eq. (2.20), is
Hermitian (this can be seen by summing Eq. (2.20) over both $j$
and $k$, and making use of the resolution of identity in Eq.
(2.13).) One important kind of ensemble of states for which the
optimum strategy can be derived analytically are pure states with
equal a priori probabilities, ${\eta}_{j}=1/N$, which are also
{\em symmetric}\cite{Helstrom,Symmetric}.  A set of states is
symmetric if it satisfies the following conditions:
\begin{eqnarray}
|{\psi}_{j}{\rangle}&=&U|{\psi}_{j-1}{\rangle}=U^{j-1}|{\psi}_{1}{\rangle},
\\ * U|{\psi}_{N}{\rangle}&=&|{\psi}_{1}{\rangle},
\end{eqnarray}
for some unitary operator $U$.  We see that $U$ transforms each
state into its successor, and the final state back to the initial
state.  The optimum measurement for these states is the so-called
{\em square-root measurement}.  We define the operator:
\begin{equation}
{\Phi}=\sum_{j=1}^{N}|{\psi}_{j}{\rangle}{\langle}{\psi}_{j}|.
\end{equation}
The optimum detection operators ${\Pi}_{j}$ are of the form
\begin{equation}
{\Pi}_{j}=|{\omega}_{j}{\rangle}{\langle}{\omega}_{j}|,
\end{equation}
where the, in general unnormalised, and non-orthogonal states
$|{\omega}_{j}{\rangle}$ are given by
\begin{equation}
|{\omega}_{j}{\rangle}={\Phi}^{-1/2}|{\psi}_{j}{\rangle}.
\end{equation}
It is because of the presence of ${\Phi}^{-1/2}$ on the right hand
side of this expression that this measurement is known as the
square-root measurement.  For equally-probable symmetric states,
this measurement attains the minimum error probability
\begin{equation}
P_{E}({\mathrm
opt})=1-{\frac{1}{N}}\sum_{j=1}^{N}|{\langle}{\psi}_{j}|{\Phi}^{-1/2}|{\psi}_{j}{\rangle}|^{2}.
\end{equation}
The simplest set of symmetric states is that of just two states,
which we examined above. Looking back at Eq. (2.7), we see that if
the orthogonal basis states $|{\pm}{\rangle}$ are to be regarded
as the spin up/down states of a spin-1/2 particle with respect to
the {\em z}-axis, then the {\em z} component of the spin vector,
${\sigma}_{z}$, acts on these states to give
${\sigma}_{z}|{\pm}{\rangle}={\mp}|{\mp}{\rangle}$. Using this
property, it can easily be seen that
${\sigma}_{z}|{\psi}_{\pm}{\rangle}=|{\psi}_{\mp}{\rangle}$. This
operator is unitary and satisfies ${\sigma}_{z}^{2}=1$, so these
states satisfy the symmetric states conditions in Eqs.
(2.23-2.24).

The next simplest case is that of three states.  Three symmetric
photon polarisation states are
\begin{eqnarray}
|{\psi}_{1}{\rangle}&=&|{\leftrightarrow}{\rangle}, \\ *
|{\psi}_{2}{\rangle}&=&\frac{-|{\leftrightarrow}{\rangle}+{\sqrt{3}}|{\updownarrow}{\rangle}}{2},\\
*
|{\psi}_{3}{\rangle}&=&\frac{-(|{\leftrightarrow}{\rangle}+{\sqrt{3}}|{\updownarrow}{\rangle})}{2}.
\end{eqnarray}
These states are illustrated schematically in Figure 5, where we
see that they are distributed around a circle, with equal angular
spacing of $2{\pi}/3$ radians.  This ensemble of states is
sometimes called the {\em trine} ensemble.  If these states have
equal a priori probabilities, then the minimum error probability
is equal to 1/3, and the optimum strategy is given by the
detection operators
\begin{equation}
{\Pi}_{j}=\frac{2}{3}|{\psi}_{j}{\rangle}{\langle}{\psi}_{j}|.
\end{equation}
The first method of carrying out the optimum state discrimination
measurement for a trine ensemble of photon polarisation states was
proposed by Sasaki {\em et al}\cite{Sasaki}.  Phoenix {\em et
al}\cite{Trine} explored the potential of this ensemble of states
in quantum cryptography, showing that, for a certain, novel three
state key distribution protocol, it is the trine ensemble which
can be used to generate secret key bits most efficiently.

The square-root measurement for three symmetrical photon
polarisation states can be performed with current technology.
Indeed, at the time of writing, experimental demonstrations of
both this measurement, and also the more complex minimum error
probability discrimination of four non-coplanar states arranged as
a tetrahedron, have just been carried out by Clarke {\em et
al}\cite{Clarke1}.
\section{Unambiguous state discrimination}
\renewcommand{\theequation}{3.\arabic{equation}}
\setcounter{equation}{0} \subsection{Error-free discrimination
between two non-orthogonal states.}
As we have seen, the formalism
of generalised measurements offers greater scope for the
possibility of discriminating between non-orthogonal quantum
states than simple von Neumann measurements.  One of the main
advantages conferred by generalised measurements is the fact that
the number of distinguishable outcomes can be arbitrarily large.
The number of outcomes possible with a von Neumann measurement is
restricted to be no greater than the number of dimensions of the
system's state space.  This means that quantum hypothesis testing,
with a full set of outcomes corresponding to each of the states,
is not generally possible with von Neumann measurements, in
particular, if the states are linearly dependent. As we saw in the
case of the trine ensemble, it is necessary under such
circumstances to use a generalised measurement.

Being able to perform measurements with an arbitrary number of
outcomes, we might ask ourselves if there is anything to be gained
if not every outcome need correspond to the detection one of the
states.  In other words, do we gain anything if we allow our
measurement to have inconclusive results?  This issue was first
examined in 1987 by Ivanovic\cite{Ivanovic}, who made the
startling discovery that the possibility of occasionally obtaining
inconclusive results permits {\em error free} discrimination
between non-orthogonal states. Ivanovic showed that when the
result of this measurement is not inconclusive, it is always
correct.

To see how this may be done, consider again the two states
$|{\psi}_{\pm}{\rangle}$ defined in Eq. (2.7).  Let us now
introduce the additional states
\begin{equation}
|{\psi}_{\pm}^{\perp}{\rangle}={\sin}{\theta}|+{\rangle}{\pm}{\cos}{\theta}|+{\rangle}.
\end{equation}
Notice that $|{\psi}_{+}^{\perp}{\rangle}$ is orthogonal to
$|{\psi}_{-}{\rangle}$, and likewise with
$|{\psi}_{-}^{\perp}{\rangle}$ and $|{\psi}_{+}{\rangle}$.
Consider now a generalised measurement described by the following
detection operators:
\begin{equation}
{\Pi}_{\pm}=\frac{P_{\pm}}{|{\langle}{\psi}_{\pm}^{\perp}|{\psi}_{\pm}{\rangle}|^{2}}|{\psi}_{\pm}^{\perp}{\rangle}{\langle}{\psi}_{\pm}^{\perp}|,\;\;\;\;\;\;{\Pi}_{?}=1-{\Pi}_{+}-{\Pi}_{-}.
\end{equation}
The meaning of the coefficients $P_{\pm}$ will become apparent
shortly. Since $|{\psi}_{+}^{\perp}{\rangle}$ is orthogonal to
$|{\psi}_{-}{\rangle}$, we see that
${\langle}{\psi}_{-}|{\Pi}_{+}|{\psi}_{-}{\rangle}=0$, and so the
probability of obtaining the result `+' for this state must be
zero.  Likewise,
${\langle}{\psi}_{+}|{\Pi}_{-}|{\psi}_{+}{\rangle}=0$, so we will
never obtain the result `$-$' for the initial state
$|{\psi}_{+}{\rangle}$. Thus, whenever we obtain one of these two
results, we can retrodict {\em exactly} what the initial state
was.  We can easily see that
\begin{equation}
{\langle}{\psi}_{\pm}|{\Pi}_{\pm}|{\psi}_{\pm}{\rangle}=P_{\pm}.
\end{equation}
This means that $P_{\pm}$ is the probability, given that the
system was prepared in the state $|{\psi}_{\pm}{\rangle}$, that
this state will be identified unambiguously.

Unless the states are orthogonal, these probabilities cannot attain the va1ue of 1.  There
is a third result, the inconclusive result `?', the probability of which for each state is
equal to the expectation value of the operator ${\Pi}_{?}$.  As with the quantum
hypothesis testing strategy we discussed in the preceding section, it is important to
optimise this measurement.  This means obtaining the maximum unambiguous discrimination
probability, or equivalently, the minimum probability of inconclusive results. To obtain
this, we must know the a priori probabilities ${\eta}_{\pm}$ of the two states. Given
these, the total probability $P_{?}$ of obtaining an inconclusive result is
\begin{equation}
P_{?}=1-\sum_{j=+,-}{\eta}_{j}P_{j}.
\end{equation}
The variational problem whose solution is $P_{?}({\mathrm opt})$
essentially consists of determining the values of $P_{\pm}$ which
minimise Eq. (3.4) subject to the constraint that the operator
${\Pi}_{?}$ is positive. For two states with equal a priori
probabilities ${\eta}_{\pm}=1/2$, it was established through the
work of Ivanovic\cite{Ivanovic}, Dieks\cite{Dieks1} and
Peres\cite{Peres} that the minimum attainable value of the
inconclusive result probability is given by
\begin{equation}
P_{?}({\mathrm opt})=|{\langle}{\psi}_{+}|{\psi}_{-}{\rangle}|.
\end{equation}
This Ivanovic-Dieks-Peres (IDP) limit is obtained when $P_{+}$ and
$P_{-}$ are both equal to
$1-|{\langle}{\psi}_{+}|{\psi}_{-}{\rangle}|$, which is also the
total probability of obtaining a conclusive, correct result.   A
more general bound was later obtained by Jaeger and
Shimony\cite{JShimony}, which solves the problem for unequal a
priori probabilities.  A particularly illuminating discussion of
the Jaeger-Shimony result in the context of quantum communications
was given by Ban\cite{Ban}

It is important to understand what happens to the state of the
system when an inconclusive result is obtained. It might be
tempting to imagine that this outcome is of little importance,
since we could repeat the measurement. Unfortunately, this is not
the case.  If an inconclusive result is obtained, then the states
$|{\psi}_{\pm}{\rangle}$ undergo a transformation. In general, for
a fixed pair of states, the lower $P_{?}$ is for the measurement,
then the closer to each other, in terms of their overlap, the
possible states will be after the transformation.  As $P_{?}$
reaches $P_{?}({\mathrm opt})$ in Eq. (3.5), both states are
transformed into the same state, rendering any further attempt to
discriminate between them futile. Unambiguous discrimination can
then be regarded as a kind of gambling operation.  The states
$|{\psi}_{\pm}{\rangle}$ are distinguishable to some extent,
though not completely so.  We can gamble this partial
distinguishability in the hope of obtaining complete
distinguishability, and will succeed with probability $1-P_{?}$.
If we lose, however, the states become less distinguishable than
they were initially.

As with the measurements described in the preceding section,
photon polarisation states are ideally suited to the experimental
realisation of this measurement\cite{Huttner}. The two states can
be represented as
\begin{equation}
|{\psi}_{\pm}{\rangle}={\cos}{\theta}|{\updownarrow}{\rangle}{\pm}{\sin}{\theta}|{\leftrightarrow}{\rangle}.
\end{equation}
where again, $0{\le}{\theta}{\le}{\pi}/4$.  Consider now the
interferometric setup depicted in Figure 6. A photon prepared in
one of these states enters polarising beamsplitter PBS1. This is
oriented so as to transmit photons which are horizontally
polarised, and reflect the vertically polarised ones.  The
vertical polarisation component travels up to the ordinary
beamsplitter BS, which has a transmission coefficient
\begin{equation}
t=\frac{\sqrt{{\cos}2{\theta}}}{{\cos}{\theta}}.
\end{equation}
If a photon is transmitted here, then it will result in a count at
detector $D_{?}$, and give an inconclusive result.  The
probability of this occurrence is the product of the probability
that the photon is vertically polarised (to enable it to travel
along the upper branch of the interferometer), which, from Eq.
(3.6), is given by ${\cos}^{2}{\theta}$, and the probability of
transmission, given by $t^{2}$. The result is simply
${\cos}2{\theta}=|{\langle}{\psi}_{+}|{\psi}_{-}{\rangle}|$, which
is the minimum probability in Eq. (3.5).

An inconclusive result is not obtained if this component is
reflected at BS, in which case it will encounter a second
polarising beamsplitter PBS2, which, like PBS1, transmits
horizontal and reflects vertical polarisation. In the absence of a
detection at PBS1, the photon will emerge from PBS2, its state
having undergone the transformation
\begin{equation}
|{\psi}_{\pm}{\rangle}{\rightarrow}\frac{1}{\sqrt
2}(|{\updownarrow}{\rangle}{\pm}|{\leftrightarrow}{\rangle}).
\end{equation}
These states are orthogonal, and can be distinguished using a
third polarising beamsplitter PBS3.  As in the Barnett-Riis
experiment discussed in the preceding section, a polarising
beamsplitter, here PBS3, oriented at ${\pi}/4$ to the horizontal
can be used to deflect the photon to one of the detectors $D_{+}$
and $D_{-}$ only when its initial state was $|{\psi}_{+}{\rangle}$
or $|{\psi}_{-}{\rangle}$ respectively. The wrong path is never
taken, so that when successful, the discrimination attempt will
always give the correct answer.

While technically feasible, the experimental apparatus shown in
Figure 6 would have to be stabilised and aligned with extreme
accuracy.  This difficulty was overcome in an ingenious variation
of this experiment reported in 1996 by Huttner {\em et
al}\cite{Huttner}.  An interferometer, such as that in Figure 6,
or other device where photons with different polarisations travel
along different paths, might seem to be essential for this type of
experiment, since the non-orthogonal initial states are
transformed into orthogonal ones using polarisation-dependent
losses (PDL).  What Huttner and collaborators realised was that
the same effect can be achieved with all photons travelling along
the same path if the medium through which they propagate itself
has PDL. Using an optical fiber with this property, they performed
the experiment using highly attenuated optical pulses
(${\approx}0.1$ photons per pulse).  For ${\theta}={\pi}/6$, they
obtained an error rate of $1.7\%$.  Comparing this with the
minimum error probability in the Helstrom measurement, which is
approximately $6.7\%$, this measurement shows a clear improvement.

One disadvantage of using a PDL fiber is the fact that photons
lost do not go to a detector, they simply do not register
anywhere.  Thus, the occasions when photons fail to result in a
click at either $D_{+}$ or $D_{-}$, which we would like to
interpret as being caused by inconclusive results, cannot be
distinguished from those null events due to the weakness of the
pulse.  However, this does not detract from the fact that the
error probability they obtained for detected photons was
significantly less than the Helstrom bound.

Nevertheless, it is important to distinguish both kinds of null
result.  At the time of writing, an experimental realisation of
the IDP measurement which follows the scheme shown in Figure 6
more closely, and produces the optimal theoretical proportions of
conclusive and inconclusive results, has just been carried out by
Clarke {\em et al}\cite{Clarke2}.
\subsection{Unambiguous discrimination between linearly
independent states}

A naturally intriguing question is: how can this type of
measurement be generalised to more than two states? We would then
consider a set of $N$ quantum states $|{\psi}_{j}{\rangle}$, with
$j=1,{\ldots},N$. Correspondingly, we would have detection
operators ${\Pi}_{j}$ satisfying
\begin{equation}
{\langle}{\psi}_{j'}|{\Pi}_{j}|{\psi}_{j'}{\rangle}=P_{j}{\delta}_{jj'}.
\end{equation}
This condition says that outcome $j$ can only occur when the
initial state is $|{\psi}_{j}{\rangle}$.  The conditional
probability, given that the system was prepared in this state,
that it will be successfully identified, is $P_{j}$.  There will
also be a further detection operator ${\Pi}_{?}$ corresponding to
inconclusive results.

In \cite{Melinear}, I showed that such a measurement strategy can
only exist if the states $|{\psi}_{j}{\rangle}$ are {\em linearly
independent}.  For $N$ linearly independent states, the form of
the detection operator ${\Pi}_{j}$ is an immediate generalisation
of that for the two state case, given by Eq. (3.2):
\begin{equation}
{\Pi}_{j}=\frac{P_{j}}{|{\langle}{\psi}_{j}^{\perp}|{\psi}_{j}{\rangle}|^{2}}|{\psi}_{j}^{\perp}{\rangle}{\langle}{\psi}_{j}^{\perp}|,\;\;\;\;\;\;{\Pi}_{?}=1-\sum_{j=1}^{N}{\Pi}_{j}.
\end{equation}
The normalised state $|{\psi}_{j}^{\perp}{\rangle}$ is defined as
that which is orthogonal to all $|{\psi}_{j'}{\rangle}$  for
$j{\neq}j'$.  Up to an overall phase, this state is unique and
known as the {\em reciprocal state}. The relationship between the
reciprocal states to the original states $|{\psi}_{j}{\rangle}$ is
exactly analogous to that in crystallography between the
reciprocal vectors and the Bravais lattice vectors, where each
member of the former set of vectors is orthogonal to all but one
member of the latter set\cite{Guinier}.

As we saw in the case of unambiguous discrimination between just
two states, it is important to examine how the possible initial
states are transformed when an inconclusive result is obtained.
Optimal unambiguous discrimination measurements on a pair of
states transforms them into the same state if the measurement
fails.  It is shown in\cite{Melinear} that for a set of $N$
linearly independent states, a failure will transform them into a
linearly dependent set, making any further attempt to discriminate
between them without errors impossible.

Nevertheless, an inconclusive result does not necessarily erase
all of the information about the state (except in the case $N=2$),
and it is still possible to obtain some information.  If the
unambiguous discrimination attempt fails, one can still carry out
the quantum hypothesis testing strategy, described in the
preceding section, on the resulting linearly dependent states.

An important question is whether or not one can obtain an analytic
expression for the minimum probability of inconclusive results for
more than two states.  By analogy with the two state case, we take
the state $|{\psi}_{j}{\rangle}$ to have a priori probability
${\eta}_{j}$, and see that the total probability of inconclusive
results is given by
\begin{equation}
P_{?}=1-\sum_{j=1}^{N}{\eta}_{j}P_{j}.
\end{equation}
As in the two state case, the variational problem consists of determining the $P_{j}$
which minimise $P_{?}$ subject to the constraint that the inconclusive result operator
${\Pi}_{?}$ is positive.

Like the quantum hypothesis testing strategy, it is difficult to find explicit solutions
for this optimisation problem for unambiguous discrimination with arbitrary states.
However, also as with the hypothesis testing strategy described in the preceding section,
the problem is explicitly soluble for equally probable symmetrical states, which satisfy
Eqs. (2.23-2.24)\cite{Mesymmetric}. In fact, at the time of writing, this is the only
known solution for more than two states.

This solution is expressed in terms of a special representation of
these states. Here, of course, we are concerned with symmetrical
states which are also linearly independent. Note that these are
completely distinct conditions.  The trine ensemble of three
symmetric states of a two-level system, discussed in the preceding
section, is clearly linearly dependent. If the
$|{\psi}_{j}{\rangle}$ are both linearly independent and
symmetric, then they may be written as
\begin{equation}
|{\psi}_{j}{\rangle}=\sum_{k=1}^{N}c_{k}{\exp}\left(\frac{2{\pi}ijk}{N}\right)|k{\rangle}.
\end{equation}
for some coefficients $c_{k}$ and orthonormal states
$|k{\rangle}$.  In fact, the $|k{\rangle}$ are the eigenstates of
the unitary operator $U$ in Eqs. (2.23) and (2.24).   The
coefficients $c_{k}$ satisfy the normalisation condition
$\sum_{k=1}^{N}|c_{k}|^{2}=1$. If the $|{\psi}_{j}{\rangle}$ have
equal a priori probabilities ${\eta}_{j}=1/N$, then the minimum
value of $P_{?}$ is given by
\begin{equation}
P_{?}({\mathrm opt})=N{\times}\min_{k}|c_{k}|^{2}.
\end{equation}
For optimum unambiguous discrimination between linearly
independent symmetric states, it turns out that the conditional
probabilities $P_{j}$ are equal for all states, and therefore,
from the equality of the a priori probabilities that we have been
assuming, equal to $1-P_{?}({\mathrm opt})$.  This measurement has
been found to have some novel applications.  For example, it has
been shown by Dusek {\em et al}\cite{Norbert} that the possibility
of such a measurement has worrying implications for quantum
cryptography. These authors showed that for realistic
implementations of the first quantum key distribution protocol,
devised by Bennett and Brassard in 1984 (BB84), the use of this
measurement as an eavesdropping strategy can render the protocol
insecure for certain detector efficiencies.

The optimisation problem for more than two states has also been
examined by Peres and Terno\cite{Terno1}.  These authors gave a
particularly detailed examination of the geometry and topology of
the set of detection operators for 3 states, and showed how their
method can be generalised to an arbitrary number of states.

In the following section, we shall look at a further interesting
application of unambiguous discrimination between symmetric
states, which relates to the manipulation of quantum entanglement.
\section{State discrimination and entanglement}
\renewcommand{\theequation}{4.\arabic{equation}}
\setcounter{equation}{0} \subsection{Entanglement and quantum
correlations}
 Recently, much attention has been paid to a
peculiar type of correlation between quantum systems known as {\em
entanglement}. In this section, we shall examine some of the main
properties of entanglement, placing particular emphasis on those
which are related to state discrimination.  We shall begin by
describing the type of nonlocal correlations which can occur
between systems which are entangled and show how, if it were
possible to discriminate between arbitrary quantum states, then
this could be used to transmit information across large distances
instantaneously, in violation of the special theory of relativity.

Entanglement is produced  when quantum systems interact with one
another.  If an operation on a pair of quantum systems involves no
interaction between them, then it may be implemented as a series
of distinct operations on the individual component systems,
otherwise known as {\em local quantum operations}, perhaps
together with classical communication between the agencies in
possession of the components.

It is widely acknowledged that the fundamental properties of
entanglement are that it is invariant under local {\em unitary}
quantum operations and cannot, on average, increase under
arbitrary local quantum operations and classical
communication\cite{Plenio,VedralPRL}. However, if we have a state
which is slightly entangled, it is possible to act only on one of
the subsystems in a way which {\em sometimes} produces more
entanglement. Although using only local quantum operations, and
possibly classical communication, we cannot increase entanglement
on average, Bennett {\em et al}\cite{Bennett1} discovered that we
can gamble a small amount of initial entanglement with the
possibility of obtaining more.  This idea of obtaining, with some
probability, a gain which cannot be acquired deterministically
also lies at the heart of unambiguous discrimination.  We shall
see that the relationship between these two operations is far from
superficial\cite{Melinear,Meent}. In fact, the local operation on
one of the entangled subsystems which transforms the entire state
into a {\em maximally entangled } state with maximum probability
is also the operation which performs optimal unambiguous
discrimination between a related set of symmetrical
states\cite{Melinear}, which we discussed in the preceding
section.

It is helpful to begin by explaining what an entangled state is.
Here, we shall consider only pure states.  Suppose that Alice and
Bob possess two quantum systems, $A$ and $B$. If these systems
have been prepared independently in the states
$|{\psi}^{1}{\rangle}$ and $|{\psi}^{2}{\rangle}$ respectively,
then the state of the combined system will be of the form
\begin{equation}
|{\psi}{\rangle}=|{\psi}^{1}{\rangle}_{A}|{\psi}^{2}{\rangle}_{B}.
\end{equation}
Such a state is known as a {\em product state}.  The significance
of this form becomes apparent when we calculate expectation values
of physical observables.  Let ${\alpha}$ be an operator observable
for $A$, and ${\beta}$ be one for $B$. Then the expectation value
of the product ${\alpha}{\beta}$ is simply
\begin{equation}
{\langle}{\psi}|{\alpha}{\beta}|{\psi}{\rangle}={\langle}{\psi}^{1}|{\alpha}|{\psi}^{1}{\rangle}{\langle}{\psi}^{2}|{\beta}|{\psi}^{2}{\rangle},
\end{equation}
that is, it is simply the product of the expectation values of the
two observables.  If these two operators are projection operators,
then they represent propositions, and their expectation values are
the probabilities that these propositions are true.  The product
${\alpha}{\beta}$, which is a projection operator on the space of
the combined system, represents the logical `and' of these two
propositions.  We see then from Eq. (4.2) that the probability of
`${\alpha}$ and ${\beta}$' being true is simply the product of the
probabilities of ${\alpha}$ being true and ${\beta}$ being true.
This implies that the truth probabilities of these two
propositions are uncorrelated. Since these propositions are
completely arbitrary, no property of particle $A$ has any
correlation with any of particle $B$.

The product state in Eq. (4.1) is not, however, the most general
type of pure state of $A$ and $B$.  The superposition principle
implies that the state of the entire system can be any linear
combination of product states such as those in Eq. (4.1).  An
entangled state is such a superposition which cannot be expressed
as a single product state. The product rule in Eq. (4.2) for
expectation values of local observables does not generally hold
for such states.   One of the most extensively studied entangled
states is the {\em singlet state}
\begin{equation}
|{\psi}{\rangle}=\frac{1}{\sqrt{2}}\left(|{u}{\rangle}_{A}|{u'}{\rangle}_{B}-|{u'}{\rangle}_{A}|{u}{\rangle}_{B}
 \right).
\end{equation}
The states $|u{\rangle}$ and $|u'{\rangle}$ are orthogonal for
each particle, and the singlet state is said to be {\em maximally
entangled}. We shall shortly examine the problem of quantifying
the amount of entanglement in general states.  For the moment
though, we shall consider only the singlet state as this can be
used to show that, if it were possible to discriminate between
arbitrary states, then information could be transmitted
instantaneously.

We begin by noting that, since $|u{\rangle}$ and $|u'{\rangle}$
are orthogonal for each system, we can construct the Hermitian
operators $U_{A}$ and $U_{B}$ which have these states as their
eigenstates.  A measurement of one of these operators can be used
to distinguish perfectly between these states.

The singlet state has an interesting symmetry property, in that
the states $|u{\rangle}$ and $|u'{\rangle}$ can be any pair of
orthogonal states at all, and $|{\psi}{\rangle}$ still has this
form shown in Eq. (4.3). We can then define another pair of
orthogonal states, $|v{\rangle}$ and $|v'{\rangle}$ for each
system, which are the eigenstates of the Hermitian operators
$V_{A}$ and $V_{B}$, and rewrite $|{\psi}{\rangle}$ as
\begin{equation}
|{\psi}{\rangle}=\frac{1}{\sqrt{2}}\left(|{v}{\rangle}_{A}|{v'}{\rangle}_{B}-|{v'}{\rangle}_{A}|{v}{\rangle}_{B}
 \right).
\end{equation}
Measurements on entangled states such as $|{\psi}{\rangle}$ enforce non-local correlations
between the subsystems.  For example, if Alice measures $U$, then the state of her
particle will collapse into either $|u{\rangle}$ or $|u'{\rangle}$.  If Bob were then to
carry out a measurement of $U$ on his system, he would obtain, {\em with unit
probability}, the opposite result, so that if Alice obtained $|u{\rangle}$, Bob would
obtain $|u'{\rangle}$, and vice versa.  The same holds true for the states $|v{\rangle}$
or $|v'{\rangle}$, or any other pair of orthogonal states. Alice is then able to predict
the result of Bob's measurement, immediately, if he measures the same observable as her.
If no signal can travel from Alice to Bob faster than the speed of light, then we might be
led to conclude that the information Alice obtains about Bob's subsequent measurement
already exists in Bob's particle.  However, since the observable is arbitrary, it would
then follow that the information describing the results of all possible measurements on
Bob's particle must already exist.  Such an interpretation would no involve superluminal
communication.  This hypothesis, known as {\em local-realism}, is in sharp contrast to the
idea of complementarity. Einstein, who discovered these correlations with Podolsky and
Rosen\cite{EPR} in 1935, expressed a preference for local realism over the alternative,
which he imagined must be related to some kind of `spooky action at a distance'.

In local-realistic theories, the apparent randomness of the
results of quantum measurements is considered to be an illusion,
perhaps due to our ignorance of some other significant parameters
or {\em hidden variables}.  It is then important to determine
whether or not the predictions of quantum mechanics can be
reproduced by a local-realistic hidden variable theory. However,
in 1964, Bell published a theorem, according to which the
correlations produced by any such theory must satisfy a certain
inequality\cite{Bell}. For suitable parameter choices, the
predictions of quantum mechanics violate Bell's inequality for all
pure entangled states, but not for product states\cite{Gisin}, and
curiously, not for some mixed entangled states\cite{Popescu}. The
predictions of quantum mechanics have generally been vindicated by
experiment, most famously in the experiments conducted by Aspect
{\em et al}\cite{Aspect}.

If these correlations are genuinely non-local, and do not result
from the relativistically causal transmission of information
between the two systems, can Alice and Bob use them to transmit
information to each other?  If Bob could discriminate, with zero
probability of error, between the four states $|u{\rangle},
|u'{\rangle}, |v{\rangle}$ or $|v'{\rangle}$, then he could tell
whether Alice measured $U_{A}$ or $V_{A}$.  If she wishes to
transmit `1' to Bob, she measures $U$. If on the other hand, she
wishes to communicate a `0' to him, she would measure $V_{A}$
instead.

The nonlocal nature of entanglement cannot be used to transmit
information in this manner, which would avoid the actually sending
of physical systems, and thus the universal speed limit ${\em c}$.
General proofs of the impossibility of superluminal communication
using entanglement and measurement have been
obtained\cite{Ghirardi}.  On the basis of the above argument,
these proofs must implicitly place restrictions on the extent to
which the state of a quantum system can be determined.

Although entanglement cannot be used for superluminal
communication, it does have several other applications. For
example, it can be used, in conjunction with a classical
communication channel, to teleport an unknown quantum state from
one location to another\cite{Teleport}. It can also be used to
transmit classical information at twice the maximum rate that can
be achieved using classical physics, using a technique known as
{\em superdense quantum coding}\cite{Coding}. Also, many of the
recently discovered advantages of using quantum systems for
computing (the best known of which is Shor's algorithm for
factorising a number in polynomial time\cite{Shor}, for which the
best known classical algorithms require exponential time) make
explicit use of entanglement to carry out computations more
efficiently than any computer operating solely by the laws of
classical physics can manage.

It is therefore important to understand the conditions under which
entanglement can be manipulated.  In particular, how do we
quantify entanglement?  If Alice and Bob share some entangled
state $|{\psi}{\rangle}$, how much entanglement does it contain?

To answer this, we have to understand the most general form of an entangled state.  We
arrived at the concept of entanglement via the superposition principle, which enabled us
to construct a linear combination of product states which is not itself a product state.
The most general {\em pure} state of a two-particle system is simply the most general
superposition of product states.  For a pair of $N$-level quantum systems, this is
\begin{equation}
|{\psi}{\rangle}=\sum_{j,k=1}^{N}b_{jk}|{\alpha}_{j}{\rangle}_{A}|{\beta}_{k}{\rangle}_{B},
\end{equation}
where, without loss of generality, the subsystem states
$|{\alpha}_{j}{\rangle}$ and $|{\beta}_{k}{\rangle}$ are taken to
be orthogonal, so that $\sum_{j,k=1}^{N}|b_{jk}|^{2}=1$.

The $b_{jk}$ are almost completely free parameters.  They are subject only to this
normalisation constraint.  Such a large number of free parameters makes the expression in
Eq. (4.5) somewhat unwieldy. Fortunately, a simpler representation of $|{\psi}{\rangle}$
can be obtained using an important result known as the {\em Schmidt decomposition
theorem}\cite{Schmidt}. This states that there exists an orthogonal basis
$|{\alpha}'_{j}{\rangle}$ for particle $A$ and $|{\beta}'_{j}{\rangle}$ for particle $B$
such that the state $|{\psi}{\rangle}$ takes the form
\begin{equation}
|{\psi}{\rangle}=\sum_{j=1}^{N}c_{j}|{\alpha}'_{j}{\rangle}_{A}|{\beta}'_{j}{\rangle}_{B},\;\;\;\;\;\;\;\;\sum_{j=1}^{N}|c_{j}|^{2}=1,
\end{equation}
i.e. using these special bases, known as the {\em Schmidt bases},
we can write the state as a single, rather than a double sum over
product states.

For a product state, only one of the $c_{j}$ is non-zero.  If more than one is non-zero,
the state is entangled. To quantify this entanglement, we have to form the density
operator for one of the subsystems.  These are known as {\em reduced density operators},
and are denoted by ${\rho}_{A}$ and ${\rho}_{B}$.  The reduced density operator of either
subsystem is formed by taking the trace of density operator of the entire system with
respect to the other subsystem, i.e.
\begin{equation}
{\rho}_{A}={\mathrm
Tr}_{B}(|{\psi}{\rangle}{\langle}{\psi}|),\;\;\;\;\;\;{\rho}_{B}={\mathrm
Tr}_{A}(|{\psi}{\rangle}{\langle}{\psi}|).
\end{equation}
We find that these are
\begin{equation}
{\rho}_{A}=\sum_{j=1}^{N}|c_{j}|^{2}|{\alpha}'_{j}{\rangle}{\langle}{\alpha}'_{j}|,\;\;\;\;\;\;{\rho}_{B}=\sum_{j=1}^{N}|c_{j}|^{2}|{\beta}'_{j}{\rangle}{\langle}{\beta}'_{j}|.
\end{equation}
We are now in a position to quantify the amount of entanglement in
the state $|{\psi}{\rangle}$.  The entanglement, or entropy of
entanglement, $E({\psi})$, is given by the {\em von Neumann
entropy}\cite{Popescu} of either of the reduced density operators:
\begin{eqnarray}
E({\psi})&=&-{\mathrm Tr}_{A}{\rho}_{A}{\log}{\rho}_{A}={\mathrm
Tr}_{B}{\rho}_{A}{\log}{\rho}_{B} \nonumber \\ *
&=&-\sum_{j=1}^{N}|c_{j}|^{2}{\log}|c_{j}|^{2}.
\end{eqnarray}
The logarithm is conventionally taken to have base 2. Entanglement is measured in {\em
ebits}.  If only one $c_{j}$ is non-zero, corresponding to a product state, then
$E({\psi})=0$. At the other extreme, if all $|c_{j}|^{2}$ are equal, then normalisation
implies that $|c_{j}|^{2}=1/N$, and $E({\psi})={\log}N$.  Such states are called {\em
maximally entangled states} since they possess the most entanglement for a given $N$. The
singlet state in Eq. (4.3) is a maximally entangled state, and has 1 ebit of entanglement.

One of the major advantages of using the entropy of entanglement
to quantify this property is the fact that it is additive.  To
understand the meaning of this, suppose that Alice and Bob share 2
entangled states, $|{\psi}{\rangle}$ and $|{\psi}'{\rangle}$. How
much entanglement do they possess? Considering these two entangled
systems individually, we would conclude that the total
entanglement shared by Alice and Bob is just the sum of the
entanglements of $|{\psi}{\rangle}$ and $|{\psi}'{\rangle}$.  If,
however, we consider them to be composite parts of a larger
entangled system in the state
$|{\psi}_{L}{\rangle}=|{\psi}{\rangle}|{\psi}'{\rangle}$, then the
entanglement shared by Alice and Bob is that in
$|{\psi}_{L}{\rangle}$.  Clearly, to quantify shared entanglement
unambiguously, it is necessary that an entanglement measure $E$
satisfies $E({\psi}_{L})=E({\psi})+E({\psi}')$.  Fortunately, the
entropy of entanglement has this desirable additive property for
pure entangled states.  However, no entanglement measure has yet
been shown to be additive over the set of all {\em mixed}
entangled states.  Recently, though, additivity has been shown to
hold for one of the most important entanglement measures, the {\em
relative entropy of entanglement}, for a large class of
states\cite{Relative}.
\subsection{Entanglement concentration and unambiguous state
discrimination}

The entanglement $E({\psi})$ cannot be {\em deterministically}
increased by acting on the subsystems individually, even if
classical communication is allowed between Alice and Bob. However,
it is possible to use an unambiguous discrimination-type
measurement on either $A$ or $B$ which will, with some
probability, convert $|{\psi}{\rangle}$ into a maximally-entangled
state.  Such an operation is known as {\em entanglement
concentration}.  For the sake of definiteness, we let the
measurement be performed by Alice on particle $A$.  We first make
use of a new orthogonal basis set $|y_{k}{\rangle}$ for Bob's
particle.  These states are defined through
\begin{equation}
|{\beta}'_{j}{\rangle}=\frac{1}{\sqrt
N}\sum_{k=1}^{N}{\exp}\left({\frac{2{\pi}ijk}{N}}\right)|y_{k}{\rangle}.
\end{equation}
This expression allows us to rewrite the partly entangled state
$|{\psi}{\rangle}$ in Eq. (4.6) as
\begin{equation}
|{\psi}{\rangle}=\frac{1}{\sqrt
N}\sum_{k=1}^{N}|x_{k}{\rangle}_{A}|y_{k}{\rangle}_{B},
\end{equation}
where we have introduced another set of new states
$|x_{k}{\rangle}$, defined by
\begin{equation}
|x_{k}{\rangle}=\sum_{j=1}^{N}c_{j}{\exp}\left({\frac{2{\pi}ijk}{N}}\right)|{\alpha}'_{j}{\rangle}.
\end{equation}
Both the orthonormality of the $|y_{k}{\rangle}$ and the representation in Eq. (4.11)
follow from the relation
\begin{equation}
\sum_{k=1}^{N}{\exp}\left({\frac{2{\pi}i(j-j')k}{N}}\right)=N{\delta}_{jj'}.
\end{equation}
The $|x_{k}{\rangle}$ are normalised, although they are not orthogonal.  If they were,
then the expression in Eq. (4.11) for $|{\psi}{\rangle}$ would represent a
maximally-entangled state, which it cannot be since our transformation of basis is
passive, doing nothing to change the entanglement of the state.  These states are,
however, linearly independent. In fact, comparing Eq. (4.12) with Eq. (3.12), we see that
they constitute a set of linearly independent symmetric states of the kind we discussed in
the preceding section.

Looking at Eq. (4.11), we can say that what prevents
$|{\psi}{\rangle}$ from being maximally entangled is the
non-orthogonality of the $|x_{k}{\rangle}$. However, unambiguous
discrimination can be regarded as an operation which transforms
non-orthogonal states into orthogonal ones.  This was made
explicit in the experimental realisation of unambiguous
discrimination for a pair of states\cite{Huttner,Clarke2},
discussed in the preceding section. To see how this can be
exploited to transform $|{\psi}{\rangle}$, with some probability,
into a maximally-entangled state, consider the detection operators
for unambiguous discrimination between the $|x_{k}{\rangle}$.  It
follows from Eq. (3.10) that these are
${\Pi}_{k}=P_{k}|x^{\perp}_{k}{\rangle}{\langle}x^{\perp}_{k}|/|{\langle}x^{\perp}_{k}|x_{k}{\rangle}|^{2}$.
The states $|x^{\perp}_{k}{\rangle}$ are the reciprocal states
corresponding to the $|x_{k}{\rangle}$.  These operators, together
with ${\Pi}_{?}=1-\sum_{k=1}^{N}{\Pi}_{k}$, represent a
measurement whose possible outcomes are the $N$ states
$|x_{k}{\rangle}$, and the inconclusive result.  To use this type
of measurement for entanglement concentration, it is more
appropriate to consider just two outcomes, described by the
inconclusive result operator, ${\Pi}_{?}$, and the following
operator
\begin{equation}
{\Pi}_{O}=\sum_{k=1}^{N}{\Pi}_{k}.
\end{equation}
By definition, this pair of operators forms the required resolution of identity in Eq.
(2.13). The reason for subscript {\em O} in Eq. (4.14) will become apparent shortly.
This pair of detection operators describes a measurement having two outcomes: success or
failure of the state discrimination measurement.  It does not tell us which state has been
detected when it succeeds.  This might not seem very useful at first sight. However, the
state transformation generated by this measurement is precisely that which transforms
$|{\psi}{\rangle}$ into a maximally entangled state.  As we discussed in section II, to
determine how the state of a system is transformed by a generalised measurement, we need
to find an operator $A$ such that ${\Pi}=A^{\dagger}A$, when the result corresponds to the
detection operator ${\Pi}$.  For the operator ${\Pi}_{O}$, the corresponding
transformation operator is the {\em orthogonalisation operator} $A_{O}$
\begin{equation}
A_{O}=\sum_{k=1}^{N}\frac{P_{k}^{1/2}}{{\langle}x^{\perp}_{k}|x_{k}{\rangle}}|{\phi}_{k}{\rangle}{\langle}x_{k}^{\perp}|,
\end{equation}
where the states $|{\phi}_{k}{\rangle}$ may be any orthogonal basis.  The arbitrariness of
this basis is equivalent to that of the unitary transformation in Eq. (2.14).  As we saw
in the preceding section, the maximum probability of discriminating between a set of $N$
symmetric states is obtained when all of the $P_{k}$ are equal to $1-P_{?}({\mathrm
opt})$, where $P_{?}({\mathrm opt})$ is given by Eq. (3.13).

Using the prescription for state transformations in Eq. (2.15), we see that if the system
is initially prepared in one of the non-orthogonal states $|x_{k}{\rangle}$, when this
measurement succeeds, then the state will be transformed into the corresponding member of
the orthogonal basis $|{\phi}_{k}{\rangle}$. In fact, the unambiguous discrimination
measurement can be regarded as this orthogonalisation procedure, followed by a von Neumann
measurement in the orthogonal basis $|{\phi}_{k}{\rangle}$.  This is precisely what
happens in the photonic implementation of two-state discrimination we discussed in section
III.  There, the non-orthogonal photon states were, with probability $1-P_{?}({\mathrm
opt})$, transformed into the orthogonal states in Eq. (3.8) before being discriminated.

Looking at the representation we have for the entangled state
$|{\psi}{\rangle}$ in Eq. (4.11), we see that if Alice carries out
this operation on particle $A$, and if it succeeds, then the final
state will be
\begin{equation}
\frac{1}{\sqrt
N}\sum_{k=1}^{N}|{\phi}_{k}{\rangle}_{A}|y_{k}{\rangle}_{B},
\end{equation}
which is our promised maximally-entangled state.  The probability
of success for this operation is given by $1-N{\times}{\mathrm
min}|c_{j}|^{2}$.  As it happens, this is the maximum probability
of converting $|{\psi}{\rangle}$ into a maximally-entangled state.
The maximum probability of converting a non-maximally entangled
pure state of two systems into a maximally-entangled one, using
only local quantum operations and possibly classical
communication, was found by Lo and Popescu\cite{Lopop}. The
strategy we have been discussing reaches their bound.

Several additional results on transforming one entangled state
into another using only local operations and classical
communication have been found.  An important question is: under
what circumstances can one pure entangled state be transformed
into another with unit probability?  The solution to this problem
was obtained by Nielsen\cite{Nielsen}, whose results introduced
the powerful mathematical technique of majorisation to the study
of entanglement.

Deterministic transformations of this kind are not possible for all pairs of states.
Vidal\cite{Vidal} obtained a general expression for the maximum probability that any pure
entangled state can be converted into any other.
\section{Unambiguous discrimination and exact cloning}
\renewcommand{\theequation}{5.\arabic{equation}}
\setcounter{equation}{0} \subsection{Relationship between quantum
cloning and state discrimination}
 Another operation closely
related to unambiguous discrimination is exact cloning. In 1982,
it was discovered independently by Wootters and Zurek\cite{WZurek}
and Dieks\cite{Dieks2} that the state of a quantum system, if
unknown, cannot be copied.  As with discrimination, no completely
reliable procedure exists for this unless the state belongs to a
known orthogonal set.

By analogy with the possibility of unambiguous discrimination between linearly independent
states, it is possible, as was initially demonstrated by Duan and Guo\cite{Duanguo1}, to
build a machine which, with some probability, produces exact copies of such states. We
will examine the relationship between these two operations, in particular, that between
their maximum success probabilities.  In fact, for just two states, both operations can be
regarded as particular cases of a more general procedure known as {\em quantum state
separation}\cite{Mestate}, which we will also describe.

To understand the relationship between state discrimination and
cloning, suppose that Alice gives Bob one of the $N$ quantum
states $|{\psi}_{j}{\rangle}$.  He isn't told which, although
again, he knows what the possible states $|{\psi}_{j}{\rangle}$
are, and also their a priori probabilities ${\eta}_{j}$.   If he
can discriminate between them, then upon identifying the state, he
can manufacture as many further copies of it as he desires.
Therefore, if he can  discriminate between them, then he can also
clone them.

If, on the other hand, Bob could clone a set of states, then he
could also discriminate between them. This follows from the fact
that if Bob could make one copy,  he could make arbitrarily many.
He could then make use of the fact that, given a sufficiently
large number of copies of the state, he could determine the
expectation value of any observable, to an arbitrarily high degree
of accuracy, by repeatedly measuring it on the members of his
ensemble of clones. If the ensemble is large enough, he could
evaluate the expectation values of several observables, in fact,
any finite number of them, to any degree of accuracy.  If he
chooses the correct observables, it would be possible for him to
infer the state itself from the expectation values.  For an
$N$-dimensional system, the density operator is specified by
$N^{2}-1$ independent real parameters.  It is quite easy to see
why. The density operator has $N^{2}$ elements, each of which is
complex.  It is therefore determined by $2N^{2}$ real parameters.
There are $N^{2}$ constraints due to Hermiticity, and a further
constraint comes from the requirement of normalisation. Therefore,
to determine the state, Bob must know the expectation values of at
least $N^{2}-1$ observables. Fortunately, with a judicious choice
of observables, this lower bound can be attained.

The simplest example is the case of a two-level system or {\em
qubit}. The state ${\rho}$ of a qubit can be conveniently
expressed in the {\em Bloch representation}.  We are already
familiar with the representation of a qubit as a spin-1/2
particle, and the eigenstates $|{\pm}{\rangle}$ of the $z$
component of the spin, this being represented by the operator
${\sigma}_{z}$. It remains to introduce the other Cartesian
components ${\sigma}_{x}$ and ${\sigma}_{y}$ of this vector
operator, the {\em Pauli spin operator} ${\mbox{\boldmath
${\sigma}$}}$.  These component act on the eigenstates of
${\sigma}_{z}$ in the following way:
\begin{equation}
{\sigma}_{x}|{\pm}{\rangle}=|{\mp}{\rangle},\;\;\;\;\;\;\;{\sigma}_{y}|{\pm}{\rangle}={\pm}i|{\mp}{\rangle}.
\end{equation}
All three Cartesian components of ${\mbox{\boldmath ${\sigma}$}}$
are both Hermitian and unitary operators (implying that the square
of each of them is 1). They also have eigenvalues ${\pm}1$, so
that their trace is zero.  The Bloch representation of the density
operator ${\rho}$ is obtained by writing it as a combination of
these operators and the identity,
\begin{equation}
{\rho}=\frac{1}{2}(1+{\mathbf
a}.\mbox{\boldmath${\sigma}$\unboldmath}).
\end{equation}
The components of the Bloch vector ${\mathbf a}$ are real and the length of this vector
$|{\mathbf a}|$ is no greater than 1.  If it is equal to 1, then ${\rho}$ is a pure state.
If it is equal to 0, then ${\rho}=1/2$, meaning that it is a completely mixed state. Thus,
we may take the length of the Bloch vector to be an indicator of how pure the state is.

Determination of the state ${\rho}$ of a qubit clearly amounts to
finding the components of the Bloch vector, whose 3 components are
a special case of the general number of parameters $N^{2}-1$.  To
evaluate these components, Bob need only measure the expectation
values of each ${\sigma}_{k}$. These are equal to ${\mathrm
Tr}{\sigma}_{k}{\rho}=a_{k}$, where $k=x,y,z$. This can be seen
from the fact that ${\sigma}_{k}$ has zero trace, and from the
identity ${\mathrm Tr}{\sigma}_{k}{\sigma}_{l}=2{\delta}_{kl}$.

While it is useful to know that, in principle,  the expectation
values of a set of observables are sufficient to determine the
state of a system, in practice, these quantities cannot be
measured exactly, as this would require an infinite number of
measurements to be carried out.  Derka {\em et al}\cite{Knight}
showed how the state can nevertheless be estimated from the
available data using Bayesian inference techniques.

Using standard von Neumann measurements, the state can be
extracted from a large number of copies by measuring $N^{2}-1$
observables.  An interesting question is: can this number of types
of measurement be reduced if we use generalised measurements
instead? This issue was explored by Peres and Terno\cite{Terno2},
who came to the intriguing conclusion that only a {\em single}
generalised measurement is necessary.  Their strategy is
essentially as follows: consider a generalised measurement with
$N^{2}-1$ detection operators ${\Pi}_{k}$.  If we have a large
number of systems, all prepared in the same unknown state
${\rho}$, then we can determine the probability of the {\em k}th
outcome, ${\omega}_{k}$, which, by Eq. (2.11), is given by
$P({\omega}_{k}|{\rho})={\mathrm Tr}{\rho}{\Pi}_{k}$.  What Peres
and Terno realised was that there exist generalised measurements
for which the $N^{2}-1$ probabilities $P({\omega}_{k}|{\rho})$ are
one possible set of parameters which can be used to infer the
density operator itself.

From the above discussion, it is apparent that if Bob can copy the state he has, then he
can determine it, and vice versa. Unambiguous state discrimination is possible only if the
state belongs to a known, linearly independent set.  On the basis of the above argument,
we should expect that the same constraint limits the abilities of cloning machines. This
is indeed the case.  It was recently discovered by Duan and Guo\cite{Duanguo1} that only
linearly independent states can be cloned, and only with unit probability if they are
orthogonal.
\subsection{Exact cloning and unambiguous state discrimination}
An intriguing question is the following: given this symbiotic relationship between
discrimination and cloning, the essence of which is that the conditions under which one of
these operations is possible also apply to the other, are there quantitative relationships
between their optimal figures of merit?  On the basis of the relationship between
unambiguous discrimination and exact cloning, we would expect their maximum success
probabilities to be related.

We will see how to obtain a bound on the maximum probability of
cloning two equally probable states using the bound on the
probability of discriminating between two states, given by the
Ivanovic-Dieks-Peres limit in Eq. (3.5).  We then show that this
cloning bound also leads to the IDP bound.  Finally, we will show
that both bounds are special cases of a more general limit which
relates to an operation known as {\em quantum state separation}.

For two equally-probable pure states $|{\psi}_{\pm}{\rangle}$, the maximum probability of
success for unambiguous discrimination is given
$1-|{\langle}{\psi}_{+}|{\psi}_{-}{\rangle}|$, from the IDP limit in Eq. (3.5).  It
follows from this that if Bob has $M$ copies of the system, then the maximum probability
of discriminating between these $M$ copies
$|{\psi}_{\pm}{\rangle}_{1}{\ldots}|{\psi}_{\pm}{\rangle}_{M}$ is
\begin{equation}
P_{M{\infty}}({\mathrm
opt})=1-|{\langle}{\psi}_{+}|{\psi}_{-}{\rangle}|^{M}.
\end{equation}
The use of the notation $P_{M{\infty}}$ will become apparent shortly.  If this is the
maximum probability with which one can discriminate between these $M$-particle states,
then it is impossible to improve upon this bound by the following procedure.  We first
attempt an $N$ from $M$ cloning operation, that is, to transform these $M$ copies,
together with $N-M$ particles in `blank states', into $N$ copies, where  $N{\ge}M$. If
this succeeds, we then attempt to  discriminate between the $N$-particle products, which
cannot be accomplished with probability greater than
\begin{equation}
P_{N{\infty}}({\mathrm
opt})=1-|{\langle}{\psi}_{+}|{\psi}_{-}{\rangle}|^{N}.
\end{equation}
The cloning probability, which we shall write as $P_{MN}$,  must
be constrained by the fact that this compound operation cannot be
accomplished with probability greater than $P_{M\infty}({\mathrm
opt})$.  If this were not true, then $P_{M{\infty}}$ could not be
the maximum probability of distinguishing between $M$ copies of
$|{\psi}_{+}{\rangle}$ or $|{\psi}_{-}{\rangle}$.  Thus,
$P_{M\infty}({\mathrm opt}){\ge}P_{MN}P_{N\infty}({\mathrm opt})$.
In fact, it was shown in \cite{Mestate} that the equality here can
be attained, implying that the maximum cloning probability is
\begin{equation}
P_{MN}({\mathrm opt})=
\frac{1-|{\langle}{\psi}_{+}|{\psi}_{-}{\rangle}|^{M}}{1-|{\langle}{\psi}_{+}|{\psi}_{-}{\rangle}|^{N}}.
\end{equation}
This generalises an earlier result by Duan and Guo\cite{Duanguo2} that the maximum
probability of making two copies of the state given one initially is
\begin{equation}
P_{12}({\mathrm
opt})=\frac{1}{1+|{\langle}{\psi}_{+}|{\psi}_{-}{\rangle}|}.
\end{equation}
Exact cloning in this manner has not yet been realised in the
laboratory, although a quantum-computational network which
achieves this task has been proposed\cite{Meclone}.  A further
recent development in the study of probabilistic cloning machines
is that it is possible for the actual {\em number} of copies to be
a quantum, rather than a classical variable.  Pati \cite{Pati} has
shown how one can construct a `novel' cloning machine which, with
some probability, will generate a superposition of various numbers
of exact copies.

We see that the bound on the maximum probability of unambiguous discrimination implies a
corresponding bound on the maximum probability of exact cloning. As it happens, the bound
on exact cloning $P_{MN}({\mathrm opt})$ also implies that $P_{1{\infty}}({\mathrm opt})$
in Eq. (5.3) is actually the maximum probability of unambiguously discriminating between
the states $|{\psi}_{\pm}{\rangle}$.

Given one initial copy of the state, the maximum probability that
we can make $N$ copies is given by $P_{1N}({\mathrm opt})$.  We
can see from Eq. (5.5) that as $N{\rightarrow}{\infty}$,
$P_{1N}({\mathrm
opt}){\rightarrow}1-|{\langle}{\psi}_{+}|{\psi}_{-}{\rangle}|$
from above.  In this limit, the state could be inferred through
the statistics of appropriate measurements on the copies, so we
have shown how Eq. (5.5) implies that the states can be
discriminated unambiguously with probability
$P_{1{\infty}}({\mathrm opt})$. Consistency with the cloning bound
implies that no greater value than $P_{1{\infty}}({\mathrm opt})$
can be attained.  If state discrimination could be accomplished
with probability higher than $P_{1{\infty}}({\mathrm opt})$, then
we could, with the same probability, make an arbitrarily large
number of copies of the state given one initial realisation.  If
this probability was greater than $P_{1{\infty}}({\mathrm opt})$,
it would also exceed $P_{1N}$ for sufficiently large $N$.
Therefore, the discrimination bound can also be obtained from the
cloning bound.
\subsection{Quantum state separation}
We will conclude this section with a brief discussion of a general
quantum operation which has unambiguous discrimination and exact
cloning as special cases. This operation is known as {\em quantum
state separation}.

Consider two non-orthogonal quantum states
$|{\phi}^{1}_{\pm}{\rangle}$.  We would like to know the maximum
probability with which these can be transformed into another pair
of quantum states $|{\phi}^{2}_{\pm}{\rangle}$ such that
\begin{equation}
|{\langle}{\phi}^{2}_{+}|{\phi}^{2}_{-}{\rangle}|<|{\langle}{\phi}^{1}_{+}|{\phi}^{1}_{-}{\rangle}|.
\end{equation}
The (modulus of the) overlap of the final states is less than that
of the initial states, hence the term state separation.  In
\cite{Mestate}, it is shown that if both states have equal a
priori probabilities,  then the maximum value $P_{S}({\mathrm
opt})$ of this separation probability is
\begin{equation}
P_{S}({\mathrm
opt})=\frac{1-|{\langle}{\phi}^{1}_{+}|{\phi}^{1}_{-}{\rangle}|}{1-|{\langle}{\phi}^{2}_{+}|{\phi}^{2}_{-}{\rangle}|}.
\end{equation}
This operation corresponds to unambiguous discrimination when the
final states $|{\phi}_{\pm}^{2}{\rangle}$ are orthogonal.  In this
case, the denominator is equal to 1, and $P_{S}(\mathrm opt)$ is
equal to the Ivanovic-Dieks-Peres limit on the probability of
conclusively distinguishing between the states
$|{\phi}^{1}_{\pm}{\rangle}$.

Suppose instead that $|{\phi}^{1}_{\pm}{\rangle}$ represents $M$
copies of the state $|{\psi}_{\pm}{\rangle}$, together with $N-M$
particles in some collective `blank' state $|{\chi}{\rangle}$. We
then have
$|{\phi}^{1}_{\pm}{\rangle}=|{\psi}^{1}_{\pm}{\rangle}_{1}{\ldots}|{\psi}^{1}_{\pm}{\rangle}_{M}|{\chi}{\rangle}$.
We take the final states $|{\phi}^{2}_{\pm}{\rangle}$ to be $N$
copies of the state $|{\psi}_{\pm}{\rangle}$, that is,
$|{\phi}^{2}_{\pm}{\rangle}=|{\psi}^{2}_{\pm}{\rangle}_{1}{\ldots}|{\psi}^{2}_{\pm}{\rangle}_{N}$.
The modulus of the overlap between the final states,
$|{\langle}{\phi}^{2}_{+}|{\phi}^{2}_{-}{\rangle}|=|{\langle}{\psi}_{+}|{\psi}_{-}{\rangle}|^{N}$,
is less than that of the corresponding initial states,
$|{\langle}{\phi}^{1}_{+}|{\phi}^{1}_{-}{\rangle}|=|{\langle}{\psi}_{+}|{\psi}_{-}{\rangle}|^{M}$.
Exact cloning is then a further example of state separation.
Substitution of these expressions into Eq. (5.8) gives the maximum
cloning probability $P_{MN}({\mathrm opt})$ in Eq. (5.5).

In our discussion of state discrimination, we saw the importance of examining what happens
to the state of the system when the operation fails. The erasure of information that takes
place there also occurs for the more general state separation operation, and therefore
also for cloning.  When an optimal state separating operation fails, the possible initial
states $|{\phi}^{1}_{\pm}{\rangle}$ are transformed into the same state, rendering any
further separation attempt impossible.
\section{Universal state estimation and cloning}
\renewcommand{\theequation}{6.\arabic{equation}}
\setcounter{equation}{0} \subsection{Estimating a completely
unknown state}
So far, we have been examining the problem of trying
to discriminate, as best as we can, between members of a known,
finite set of states. Suppose that the state is completely
unknown.  In the two level case, Alice might give Bob a qubit
prepared in the state
\begin{equation}
|{\psi}{\rangle}=a|+{\rangle}+b|-{\rangle},
\end{equation}
and Bob has no information at all about the values of the
coefficients $a$ and $b$ (except, of course, that
$|a|^{2}+|b|^{2}=1$, for normalisation). The strategies we have
examined for finite sets are not useful here.  The set of possible
states is clearly linearly dependent, so unambiguous
discrimination between them is impossible.  More generally, any
realistic detection strategy will have a finite number of
outcomes. Since we are dealing here with an infinite set of
possible states, we cannot uniquely associate one outcome with
every possible state, even if we allow for errors.  Massar and
Popescu\cite{Massar} and Derka, Bu{\u{z}}ek and Ekert\cite{Derka}
examined this problem from a different perspective, proposing a
more realistic strategy known as {\em quantum state estimation}.
The problem can be formulated as a game. Alice gives Bob $M$
copies of the state $|{\psi}{\rangle}$, and his task is to perform
a measurement with $K$ outcomes. On the basis of the outcome he
records, he will conjecture that the state was a member of some
finite set of states $|{\omega}_{k}{\rangle}$, where
$k=1,{\ldots},K$.  In general, his guess will be wrong, and the
idea is to construct the measurement such that the conjectured
state is, on average, as close as possible to the actual state.

The accuracy of Bob's guess is measured by a score function.
Slightly different score functions are chosen by the two sets of
authors, although their end results are the same. The experimental
significance of their functions can be appreciated if the qubits
are realised as photon polarisation states.  Massar and Popescu
take the score function to be ${\cos}^{2}({\alpha}/2)$, where
${\alpha}$ is the angle between the actual and guessed directions
of polarisation.  Derka {\em et al} use ${\cos}^{2}{\alpha}$. This
latter choice has a significance for general quantum systems. It
is known as the {\em fidelity}.  If the actual state of the system
is $|{\psi}{\rangle}$ and the guessed state is
$|{\omega}{\rangle}$, then the fidelity $F({\omega}|{\psi})$ is
simply the square-overlap between them,
\begin{equation}
F({\omega}|{\psi})=|{\langle}{\omega}|{\psi}{\rangle}|^{2}.
\end{equation}
The fidelity has the following interpretation.  Consider a
measurement designed to determine whether or not a quantum system
has been prepared in the state $|{\omega}{\rangle}$.  The best
measurement is a so-called {\em maximal measurement}. This is a
von Neumann measurement of an operator observable which has
$|{\omega}{\rangle}$ among its eigenstates.  If the initial state
of the system is $|{\psi}{\rangle}$, then the fidelity $F$ in Eq.
(6.2) is the probability that the result of this measurement is
`yes'. The fidelity is then a natural and practical measure of how
closely the states $|{\omega}{\rangle}$ and $|{\psi}{\rangle}$
resemble each other.

In the context of polarisation, the significance of the fidelity
can be appreciated by considering Malus' Law in its photonic form.
This tells us that the probability that the actual photon state
would pass through a polariser designed to transmit photons in the
guessed polarisation state is ${\cos}^{2}{\alpha}$, i.e., it is
equal to the fidelity.

The half-angle formula of Massar and Popescu is also useful for
the following reason:  if the fidelity score is used, then when
the real and guessed polarisations are orthogonal, we would obtain
a score of 0. However, this corresponds to as much an information
gain as in cases when the score reaches 1.  This is because a
fidelity of 0 is only obtained when the guess state
$|{\omega}{\rangle}$ and the actual state $|{\psi}{\rangle}$ are
orthogonal.  A fidelity of 1 could then be obtained by replacing
each guess state by the state orthogonal to it. However, this
property presents no problems if we are interested in maximising
the score function, averaged over all states, which we shall be.
Nevertheless, Massar and Popescu's half-angle score function
avoids this ambiguity. Its minimum value is 1/2, which corresponds
to the actual state having probability of 1/2 of being in either
the guessed state or the one orthogonal to it, which means no
information gain at all.

The best measurement for any score function is that which
maximises the average of the score over all states.  As it
happens, the maxima of both score functions are equal, although
for the sake of concreteness, we shall concentrate on the
fidelity.  The average fidelity, given $M$ initial copies of the
unknown state $|{\psi}{\rangle}$, is
\begin{equation}
\bar{F}_{M}=\sum_{k=1}^{K}\int{\cal
D}|{\psi}{\rangle}P({\omega}_{k}|{\psi})F({\omega}_{k}|{\psi}),
\end{equation}
where $P({\omega}_{k}|{\psi})$ is the probability that the guessed
state is $|{\omega}_{k}{\rangle}$ given that the actual state is
$|{\psi}{\rangle}$.  This function resembles the Bayes' cost
function in Eq. (2.4), although the positive nature of a `score'
contrasts with the negative nature of a cost, which implies that
it is desirable to maximise  Eq. (6.3), whereas we would prefer to
minimise the Bayes' cost $C_{B}$ in Eq. (2.4).  This distinction,
however, is quite superficial: it amounts merely to a difference
of sign.  There are more significant differences between these two
figures of merit.  One of the most obvious differences between
$C_{B}$ and $F_{M}$ is that the latter refers to a continuous set
of states, while the former refers to a discrete set.  Also, in
evaluating the Bayes' cost, the possible states will, in general,
have different a priori probabilities. In quantum state estimation
however, we have no a priori information about the state of the
system, so the a priori probability {\em density} is uniform.
Perhaps the most pertinent difference between the two strategies
is that, in hypothesis testing, the number of outcomes is fixed:
it is equal to the number of possible states. In contrast, there
is no a priori information about the guess states
$|{\omega}_{k}{\rangle}$ in state estimation. All properties of
the guess states, including how many of them there are, are to be
determined through the optimisation procedure, that is, the
maximisation of $\bar{F}_{M}$.

Given $M$ initial copies of the state, Derka {\em et al} showed that the maximum value of
$\bar{F}_{M}$ is
\begin{equation}
\bar{F}_{M}({\mathrm opt})=\frac{M+1}{M+2}.
\end{equation}
Massar and Popescu showed that this is also the maximum of
$\bar{F}_{M}$ if $F({\omega}_{k}|{\psi})$ is replaced by
${\cos}^{2}{\alpha}/2$.  We see that the maximum score increases
with $M$ until, in the limit as $M{\rightarrow}{\infty}$, it
attains the value of 1.

Massar and Popescu proved the {\em existence} of a finite set of
guess states which attains the optimum score.  Derka {\em et al}
provided an explicit algorithm for finding these states. Actually,
their algorithm gives the optimum measurement for any set of
states generated from some reference state $|{\psi}_{0}{\rangle}$
by a unitary, finite representation of a compact Lie group.  The
Massar-Popescu paper is concerned with the group SU(2), that is,
the group which generates all possible pure states of a qubit from
the reference state.

If Alice gives to Bob $M$ copies of the state $|{\psi}{\rangle}$,
then he will obtain the guess state $|{\omega}_{k}{\rangle}$ with
probability $P({\omega}_{k}|{\psi})$.  What state then does Bob
guess `on average'? It is simply a mixture of the guess states
weighted by their respective probabilities, which we denote by
${\rho}_{guess}$:
\begin{equation}
{\rho}_{guess}=\sum_{k=1}^{K}P({\omega}_{k}|{\psi})
|{\omega}_{k}{\rangle}{\langle}{\omega}_{k}|.
\end{equation}
An appropriate measure of how closely this average guess state resembles the actual state
${\rho}_{guess}$ is again given by the fidelity:
\begin{equation}
F_{M}({\psi})={\langle}{\psi}|{\rho}_{guess}|{\psi}{\rangle}=\sum_{k=1}^{K}P({\omega}_{k}|{\psi})F({\omega}_{k}|{\psi}).
\end{equation}
The optimum mean fidelity $\bar{F}_{M}({\mathrm opt})$ in Eq. (6.4) is just the average of
the optimum $F_{M}({\psi})$ over all states $|{\psi}{\rangle}$. The symmetry of the
optimal measurement implies that $F_{M}({\psi})$ must actually be independent of
$|{\psi}{\rangle}$. This implies that the average guess state ${\rho}_{guess}$ must have
the form:
\begin{equation}
{\rho}_{guess}=\frac{1}{2}(1-S_{M})+S_{M}|{\psi}{\rangle}{\langle}{\psi}|.
\end{equation}
where $0{\leq}S_{M}{\leq}1$.  The quantity $S_{M}$ is known as a {\em shrinking factor},
for the following reason.  Let us write the density operator
$|{\psi}{\rangle}{\langle}{\psi}|$ in the Bloch representation described in the preceding
section: that is, $|{\psi}{\rangle}{\langle}{\psi}|=(1+{\mathbf
a}.\mbox{\boldmath${\sigma}$\unboldmath})/2$, for some unit vector ${\mathbf a}$.  Then
one can show using Eq. (6.7) that ${\rho}_{guess}$ has the Bloch representation
\begin{equation}
{\rho}_{guess}=\frac{1}{2}(1+S_{M}{\mathbf
a}.\mbox{\boldmath${\sigma}$\unboldmath}).
\end{equation}
The shrinking factor $S_{M}$ decreases the length of the Bloch vector ${\mathbf a}$,
although it's direction in maintained.  If $S_{M}=1$, then ${\rho}_{guess}$ is equal to
the original state $|{\psi}{\rangle}{\langle}{\psi}|$.  On the other hand, if $S_{M}=0$,
then ${\rho}_{guess}$ is completely mixed, and contains no information about the initial
state.

The optimal state estimation strategy maximises the average
fidelity $\bar{F}_{M}$.  It also maximises the shrinking factor
$S_{M}$. This can be seen in the following way.  The shrinking
factor is independent of the actual state $|{\psi}{\rangle}$.
Therefore, $F_{M}({\psi})$ in Eq. (6.6) is simply equal to
$(1+S_{M})/2$. Substituting this into Eq. (6.3) shows that
$\bar{F}_{M}$ must have the form ${\mathrm
constant}{\times}(1+S_{M})/2$. However, this constant is unity,
due to normalisation of the integral. Thus, we have
\begin{equation}
\bar{F}_{M}({\mathrm opt})=\frac{1}{2}(1+S_{M}({\mathrm opt})),
\end{equation}
leading to
\begin{equation}
S_{M}({\mathrm opt})=\frac{M}{M+2}.
\end{equation}
If only one copy of the state is initially available, then the
maximum value of the shrinking factor is 1/3, and the Bloch vector
is reduced to 1/3 of its former length.  However, the shrinking
factor grows with increasing $M$ until, in the limit as
$M{\rightarrow}{\infty}$, the shrinking factor tends to unity.

\subsection{Universal cloning machines}
As with unambiguous discrimination, the optimal figure of merit in
Eq.  (6.4) for universal state estimation is intimately related to
the optimal efficiency of cloning.  In this case, it is the
optimal efficiency of {\em universal cloning}. The idea of a
universal quantum cloning machine (UCM) was conceived by Hillery
and Bu{\u{z}}ek\cite{Hillery}.

The idea is essentially this:  Alice gives Bob $M$ copies of a quantum system prepared in
some state $|{\psi}{\rangle}$.  All states are equally probable.  Now, we know that only
linearly independent states can be cloned exactly, so the copies produced by such a
cloning machine will necessarily be imperfect. The degree of imperfection of the clones is
most easily expressed using the Bloch representation. Universal cloning machines are
designed to copy all states equally well, and are thus of a highly symmetrical nature, If
all of the clone states are required to be identical, one of the consequences is that the
direction of the Bloch vector is identical to that of the original state. However, its
length decreases by a shrinking factor $S_{MN}$ which is independent of the state cloned
and depends only upon $M$, the number of initial, exact copies and $N$, the number of
final approximate copies.  If the initial copies are of the form shown in Eq. (5.2), then
the output ones look like
\begin{equation}
{\rho}_{out}=\frac{1}{2}(1+S_{MN}{\mathbf
a}.\mbox{\boldmath${\sigma}$\unboldmath}).
\end{equation}
Optimising a universal cloning machine means minimising the
decrease in the length of the Bloch vector.  Bruss {\em et
al}\cite{Bruss2} showed that the optimum, that is, the maximum
value of this shrinking factor is
\begin{equation}
S_{MN}({\mathrm opt})=\frac{M(N+2)}{N(M+2)}.
\end{equation}
As in universal state estimation, the performance of a UCM can be characterised by either
the shrinking factor or the fidelity between the actual state and the state obtained.  We
shall denote by $F_{MN}$ the fidelity between the output state of a UCM and the exact
state $|{\psi}{\rangle}$, given that $M$ copies of  $|{\psi}{\rangle}$ were supplied to
the UCM, which then produced $N>M$ imperfect copies.  As a consequence of symmetry,
$F_{MN}$, like the shrinking factor, is independent of $|{\psi}{\rangle}$.  From the
definition $F_{MN}={\langle}{\psi}|{\rho}_{out}|{\psi}{\rangle}$, we easily find that
$F_{MN}=(1+S_{MN})/2$, leading to
\begin{equation}
F_{MN}({\mathrm opt)}=\frac{M+N+MN}{N(M+2)}.
\end{equation}
This result had been obtained previously by Gisin and
Massar\cite{GMassar} for $1{\rightarrow}N$ cloning.  Although they
proved it to be optimal for $N{\leq}7$, they conjectured it to be
optimal for all $N$.  This conjecture was confirmed by Bruss,
Ekert and Macchiavello\cite{Bruss}.  These authors also showed
that the optimum figures of merit for universal cloning are
intimately related to the maximum fidelity obtained in universal
state estimation. The relationship between universal state
estimation and cloning is most easily expressed in terms of their
respective shrinking factors $S_{M}$ and $S_{MN}$.  As is the case
with unambiguous discrimination and exact cloning, the bounds on
the optimal figures of merit for both operations imply each other.
\subsection{Relationship between state estimation and universal
cloning}

In the remainder of this section, we will present the arguments of
Bruss {\em et al} which show that, given the optimal shrinking
factor $S_{MN}({\mathrm opt})$ for cloning, we can directly deduce
the optimum shrinking factor, $S_{M}({\mathrm opt})$, for state
estimation, and vice versa. To deduce the optimum shrinking factor
for state estimation from that on cloning, suppose that an optimum
state estimation measurement is carried out on $M$ copies of an
unknown state $|{\psi}{\rangle}$.  We can use this state
estimation procedure to make approximate copies of the state in
the following way: every time the guess state
$|{\omega}_{k}{\rangle}$ is obtained, we make $N-M$ further copies
of this guess state, for some $N>M$.  The shrinking factor for the
{\em average} guess state of each of these copies is identical to
that for optimum state estimation.  However, it cannot exceed that
for an optimum UCM: it can at most equal it, in which case the
procedure we describe would actually be an optimum UCM. This leads
to the inequality
\begin{equation}
S_{MN}({\mathrm opt}){\ge}S_{M}({\mathrm opt}),
\end{equation}
for all $N>M$.  On the other hand, suppose that given $M$ initial
copies of the state $|{\psi}{\rangle}$, we send these states to an
optimum $M{\rightarrow}N$ UCM.  The $N$ approximate copies of the
state $|{\psi}{\rangle}$ are then subjected to an optimal state
estimation measurement.  The concatenation of an $M{\rightarrow}N$
UCM and a state estimation measurement on the $N$ approximate
copies cannot lead to a higher shrinking factor than a state
estimation measurement on the $M$ original copies (for much the
same reason that a {\em probabilistic cloning machine} cannot be
used to increase the probability of {\em unambiguous}
discrimination, as we saw in the preceding section).  Bruss et al
showed that this leads to the inequality
\begin{equation}
S_{M{\infty}}({\mathrm opt}){\le}S_{M}({\mathrm opt}).
\end{equation}
Combining inequalities (6.14) and (6.15), we see that the optimal
shrinking factors for state estimation and infinite universal
cloning are equal:
\begin{equation}
S_{M{\infty}}({\mathrm opt})=S_{M}({\mathrm opt}).
\end{equation}
The optimum shrinking factor, and thus fidelity, of state
estimation can be deduced from that on universal cloning.

Let us now see how the optimal shrinking factor for cloning can be
deduced from the optimum state estimation shrinking factor.  This
follows from a natural property of an optimal UCM which we will
describe first.  Suppose that we initially have $M$ copies of the
state $|{\psi}{\rangle}$.  These are fed into an $M{\rightarrow}N$
UCM, and the corresponding shrinking factor is $S_{MN}$.  We then
feed these $N$ approximate copies to a further $N{\rightarrow}L$
UCM, for some $L>N$.  This will shrink the Bloch vector further,
by the shrinking factor $S_{NL}$.  Consider now an optimal
$M{\rightarrow}L$ UCM. If this has the shrinking factor
$S_{ML}({\mathrm opt})$, then clearly we must have
\begin{equation}
S_{MN}S_{NL}{\leq}S_{ML}({\mathrm opt}).
\end{equation}
This holds for all $L$, in particular, in the limit as $L{\rightarrow}{\infty}$.  It also
holds whether or not the $N{\rightarrow}L$ UCM is optimum, so
\begin{equation}
S_{MN}{\leq}\frac{S_{M{\infty}}({\mathrm opt})}{S_{N{\infty}}({\mathrm opt})}.
\end{equation}
To obtain an expression for the optimal shrinking factor
$S_{MN}({\mathrm opt})$, we make use of two results.  The first is
the fact that, for concatenated optimal UCMs of the kind we have
been describing, the shrinking factors multiply: that is,
inequality (6.17) becomes an equality when $S_{MN}=S_{MN}({\mathrm
opt})$.  The second is Eq. (6.16), relating the shrinking factors
for optimal infinite cloning and state estimation.  These results
imply that
\begin{equation}
S_{MN}({\mathrm opt})=\frac{S_{M}({\mathrm opt})}{S_{N}({\mathrm opt})}.
\end{equation}
Notice the formal resemblance between this relationship and that
between the maximum probabilities of unambiguous discrimination
and exact cloning in Eq. (5.5).  In fact, both arguments parallel
one another, suggesting that a deeper and more general connection
between determining the state of a quantum systems, and copying
it, could be found.

We conclude this final section by mentioning some interesting
subsequent developments related to universal state estimation and
cloning.  The results we have discussed in this section refer to
an unknown state of a 2-level system.  It is natural to enquire as
to how these results can be generalised to the case of multilevel
systems.  The generalisation of universal cloning machines to
multilevel systems has been fully worked out by Werner and
Keyl\cite{Werner,Keyl}, who gave an elegant mathematical
characterisation of the optimal such transformations.

The relationship between universal cloning and state estimation
for multilevel quantum systems has also been explored by Bruss and
Macchiavello \cite{Brmach}. For general systems, this relationship
is exactly as was shown for the 2 level case, and the same
arguments can be used to deduce the optimal figures of merit.

Although we have examined the 2-level case in some detail, the
results we have described by no means tell the whole story, even
about this simple case.  We were interested in estimating, or
copying, a state $|{\psi}{\rangle}$, given $M$ initial copies.
Suppose that the physical systems which were prepared in this
state were spin-1/2 particles.  The classical analogue of such a
system would be $M$ particles whose spins point in the same
direction.  We would thus expect that, if some of the spins were
anti-parallel to the others, the information content would be the
same, since they define the same spin axis.  Gisin and
Popescu\cite{Anti} investigated this from a quantum mechanical
point of view, and found that this is not the case, namely, that
anti-parallel spins contain more information.

One of the most exciting recent developments in the field has been
the announcement, just a few days before completion of this
article, that universal cloning of photon polarisation has been
carried out in the laboratory by Li {\em et al}\cite{CExpt}. This
experiment, together with contemporaneous demonstrations of
optimal quantum state discrimination\cite{Clarke1,Clarke2},
provide further encouragement to explore the ultimate physical
limits of information processing and transmission.
\section{Discussion}
Although the state of a quantum system is not itself an
observable, we have seen how novel measurement strategies, which
are completely consistent with the formalism of quantum mechanics,
enable one to obtain information about it.

Often these measurements are different from the standard von Neumann type discussed in
introductory quantum mechanics courses. They may instead be generalised measurements.
These typically involve the interaction of the system with another ancillary system, after
which a von Neumann measurement is carried out on the latter.  These more general
measurement strategies offer greater flexibility and scope than operations performed upon
the system of interest alone.  One of the main advantages conferred by generalised
measurements to the study of state discrimination is the fact that the number of outcomes
is not limited to the dimension of the system's state space.  In quantum hypothesis
testing, this allows us to discriminate between an arbitrarily large number of states with
some probability of obtaining a correct result.  An additional outcome can correspond to
inconclusive results, which allows linearly independent states to be unambiguously
discriminated.

We also saw how state discrimination is related to other
operations on quantum systems, such as cloning and the
manipulation of quantum entanglement.  The field of quantum
information has enjoyed rapid growth over the past few years, and
some of the most intriguing discoveries made about the
information-theoretic properties of quantum systems have been
unifications of seemingly distinct concepts such as those we have
discussed here.

\section*{Acknowledgements}

I would like to express my immense gratitude to Stephen M.
Barnett, for introducing me to this field and for our fruitful
collaborative work.  I would also like to thank him, Erling Riis
and the Journal of Modern Optics for permission to use Figure 3.
My thanks also go to Masahide Sasaki and Ozamu Hirota and his
group at Tamagawa University, Tokyo, for many enjoyable
discussions about this topic.  This work was funded by the UK
Engineering and Physical Sciences Research Council (EPSRC).

\section*{About the author}

Anthony Chefles received his BSc in physics from the University of
Strathclyde, Glasgow, in 1993, and his PhD in theoretical physics
in 1997 from the same institution. His thesis topics were quantum
nonlocality, and the quantum theory of classically-nonlinear
optical systems, in particular, those which are known to exhibit
dynamical chaos.  He gained a post-doctoral position with his
thesis supervisor, Stephen M. Barnett, and worked on problems
related to state discrimination, quantum measurement and cloning
until 1999, when he was awarded a Postdoctoral Fellowship in
Theoretical Physics by the UK Engineering and Physical Sciences
Research Council (EPSRC).  In 1999, he joined the Department of
Physical Sciences at the University of Hertfordshire.  He
continues to have an active interest in quantum state
discrimination and measurement theory, and his other current
research interests are the nonlocal properties of quantum states
and operations, quantum teleportation and quantum cryptography.

\end{document}


\title{Figures}
\input epsf
\epsfverbosetrue \maketitle
\begin{figure}[tbp]
\epsfxsize15cm
 \centerline{\epsfbox{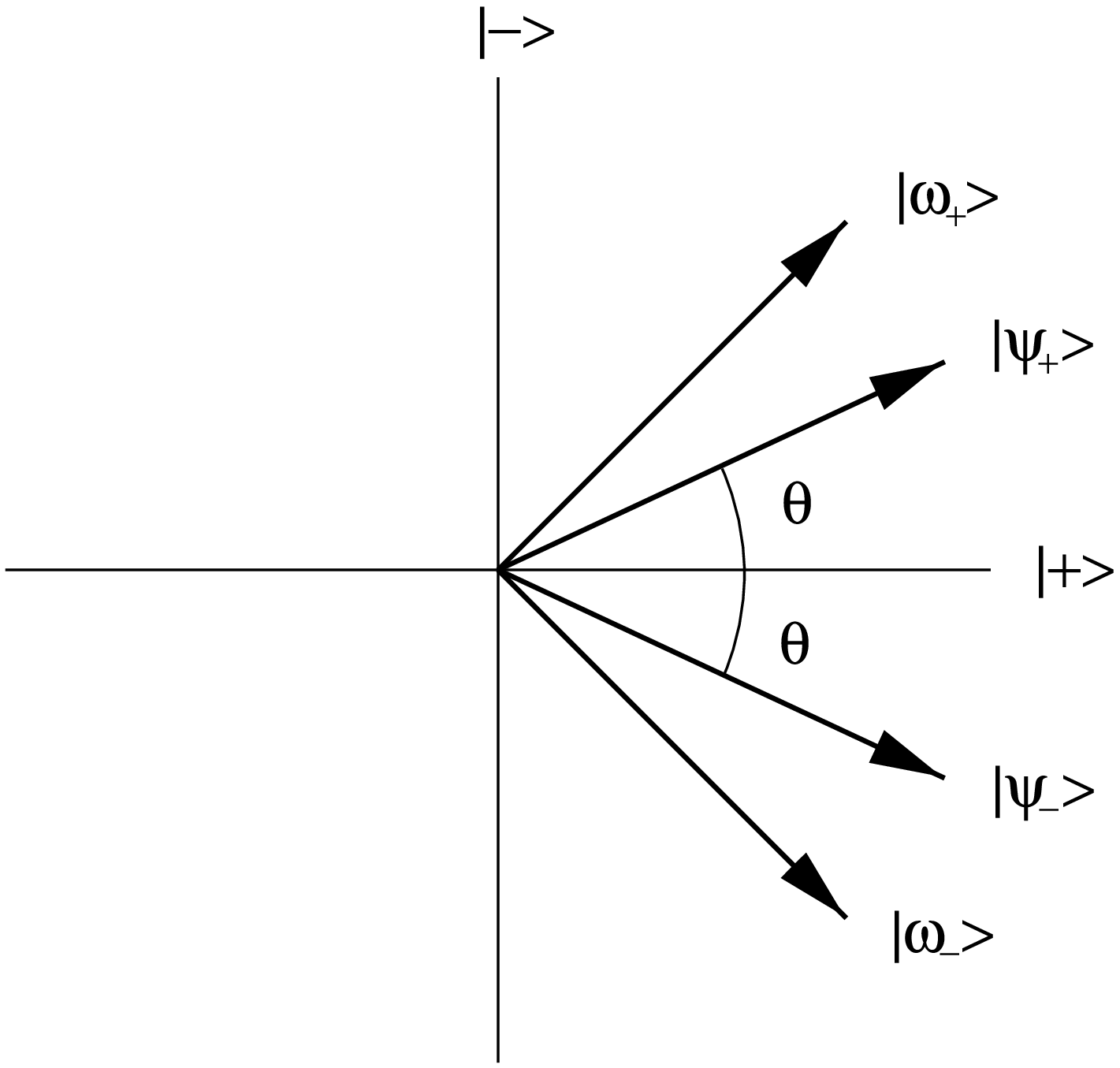}}
 \vspace{1cm}
 \caption{States involved in the Helstrom measurement. The
nonorthogonal states to be distinguished are
$\;|{\psi}_{\pm}{\rangle}$.  If these states are real-valued, then
they can be represented in a plane.  The angle can also be taken
to be acute: if the angle is obtuse, then we can rotate one of the
states by ${\pi}$ radians without altering its physical meaning,
and the new pair of states will subtend an acute angle. Any such
pair of states can be written as
$|{\psi}_{\pm}{\rangle}={\cos}{\theta}|+{\rangle}{\pm}{\sin}{\theta}|-{\rangle}$,
where $|{\pm}{\rangle}$ are some pair of orthogonal states.  If
the states $|{\psi}_{\pm}{\rangle}$ have equal a priori
probabilities, then the optimal measurement is a von Neumann
measurement in the basis
$|{\omega}_{\pm}{\rangle}=(|+{\rangle}{\pm}|-{\rangle})/{\sqrt
2}$, which gives the minimum error probability.}
\end{figure}

\newpage
\begin{figure}[tbp]
\epsfxsize15cm
 \centerline{\epsfbox{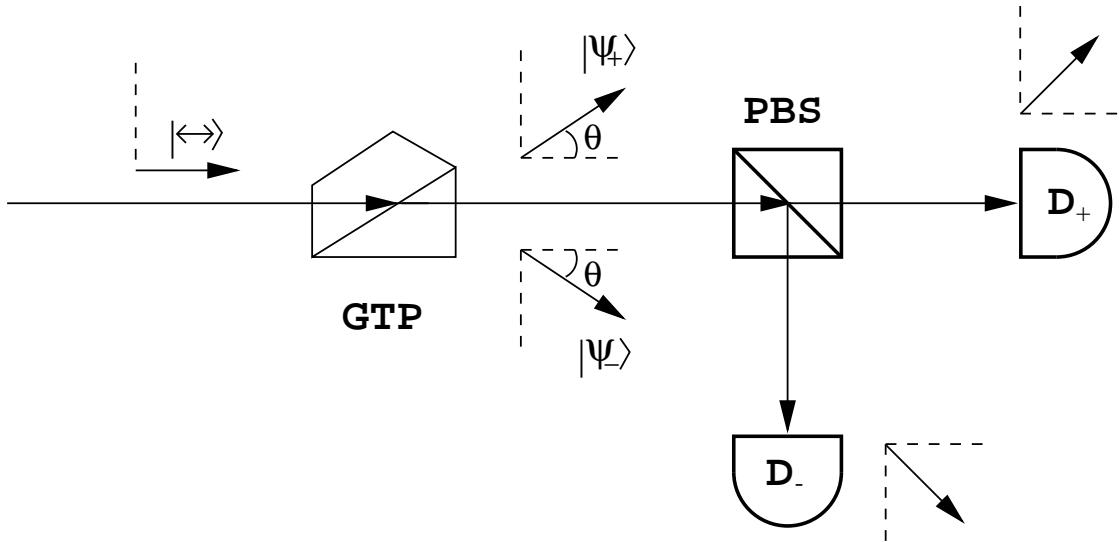}}
 \vspace{1cm}
 \caption{Depiction of the apparatus used in the
Barnett-Riis experiment.  Weak optical pulses (${\sim}0.1$
photons/pulse) were prepared in a state of horizontal polarisation
$|{\leftrightarrow}{\rangle}$.  These then encountered a
Glan-Thompson polariser.  This is a polarisation filter, whose
transmission axis was set to be in either the direction of
 $|{\psi}_{+}{\rangle}$ or $|{\psi}_{-}{\rangle}$.  The pulse
either emerged in one of these states, or was absorbed.  The
output pulses then encountered a polarising beamsplitter PBS. This
was oriented at ${\pi}/4$ radians to the horizontal, so as to
transmit the component in the direction
$(|{\leftrightarrow}{\rangle}+|{\updownarrow}{\rangle})$ to
detector $D_{+}$, and reflect the component in the orthogonal
direction
$(|{\leftrightarrow}{\rangle}-|{\updownarrow}{\rangle})$, towards
the detector $D_{-}$.  If a detection at $D_{+}$ is taken to
indicate that the initial state was $|{\psi}_{+}{\rangle}$, and
likewise with $D_{-}$ and $|{\psi}_{-}{\rangle}$, then this
measurement gives the minimum probability of error.}
\end{figure}

\newpage
\begin{figure}[tbp]
\epsfxsize15cm \centerline{\epsfbox{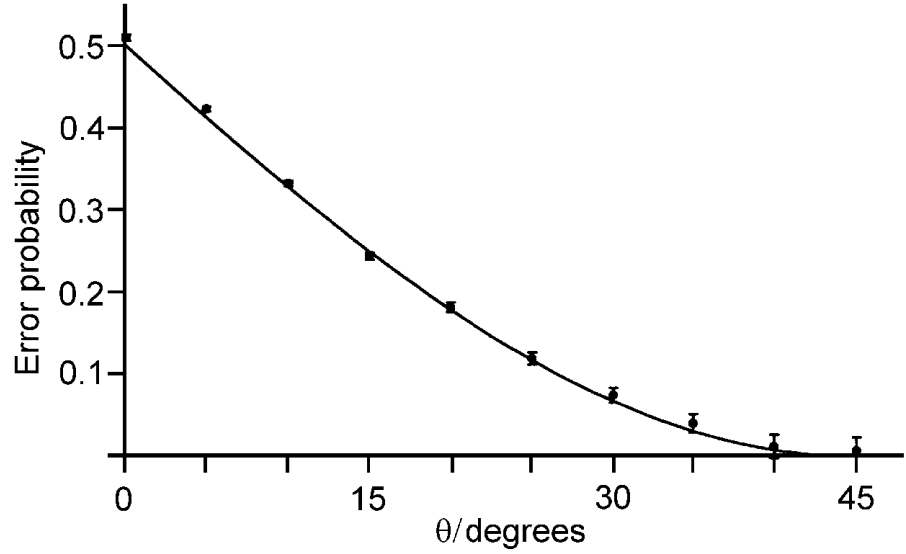}} \vspace{1cm}
\caption{Results of the Barnett-Riis experiment to discriminate
between two optical polarisation states,
$|{\psi}_{\pm}{\rangle}={\cos}{\theta}|+{\rangle}{\pm}{\sin}{\theta}|-{\rangle}$,
with minimum error probability.   The error probability is shown
versus the angle ${\theta}$.  Experimental data for specific
values of ${\theta}$ are shown in comparison with the continuous
Helstrom limit in Eq. (2.10).  Clearly the agreement between
theory and experiment is excellent.}
 \end{figure}

\newpage
\begin{figure}[tbp]
\epsfxsize15cm
 \centerline{\epsfbox{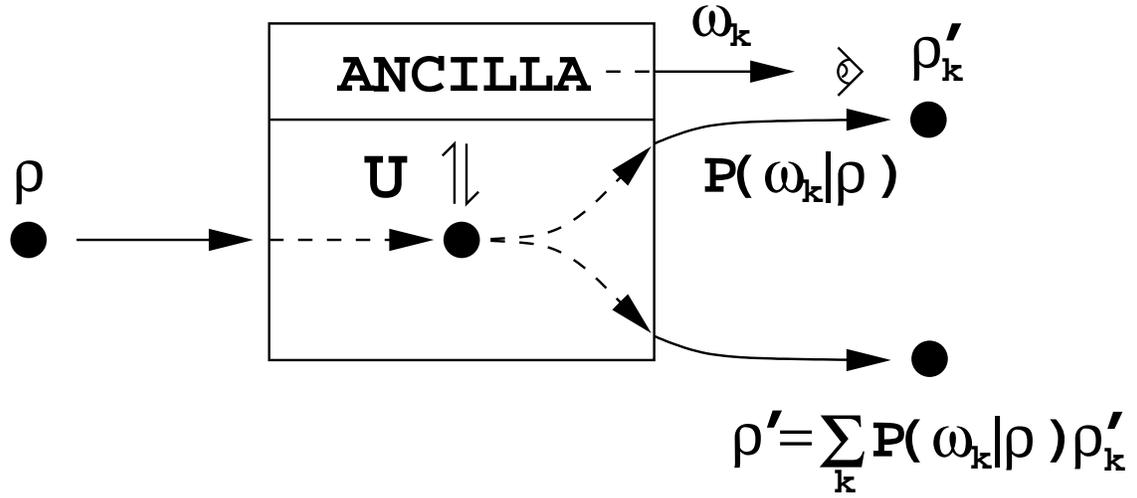}}
 \vspace{1cm}
 \caption{Diagram illustrating the general structure of
a quantum operation.  A quantum system is initially prepared in
some state, represented by the density operator ${\rho}$.  It is
then coupled to an ancilla system.  The pair of systems evolves
for some time according to some unitary operator $U$, following
which a von Neumann measurement may be performed on the ancilla.
The final state if the original system depends upon whether or not
the measurement result is recorded.  If it is, and result
${\omega}_{k}$ is obtained, then the state of the original system
undergoes the transformation
${\rho}{\rightarrow}{\rho}^{'}_{k}=A_{k}{\rho}A_{k}^{\dagger}/P({\omega}_{k}|{\rho})$,
where the $A_{k}$ are linear operators satisfying
$\sum_{k}A_{k}^{\dagger}A_{k}=1$.  Also, $P({\omega}_{k}|{\rho})$
is the probability of obtaining the result ${\omega}_{k}$, and is
given by ${\mathrm Tr}{\rho}A_{k}^{\dagger}A_{k}$.  If, on the
other hand, the measurement result is not recorded, or no
measurement is performed on the ancilla, then the final state of
the system is simply the sum of all  ${\rho}^{'}_{k}$ weighted by
their respective probabilities, that is,
${\rho}{\rightarrow}{\rho}^{'}=\sum_{k}P({\omega}_{k}|{\rho}){\rho}^{'}_{k}$.}
 \end{figure}

\newpage
\begin{figure}[tbp]
\epsfxsize15cm
 \centerline{\epsfbox{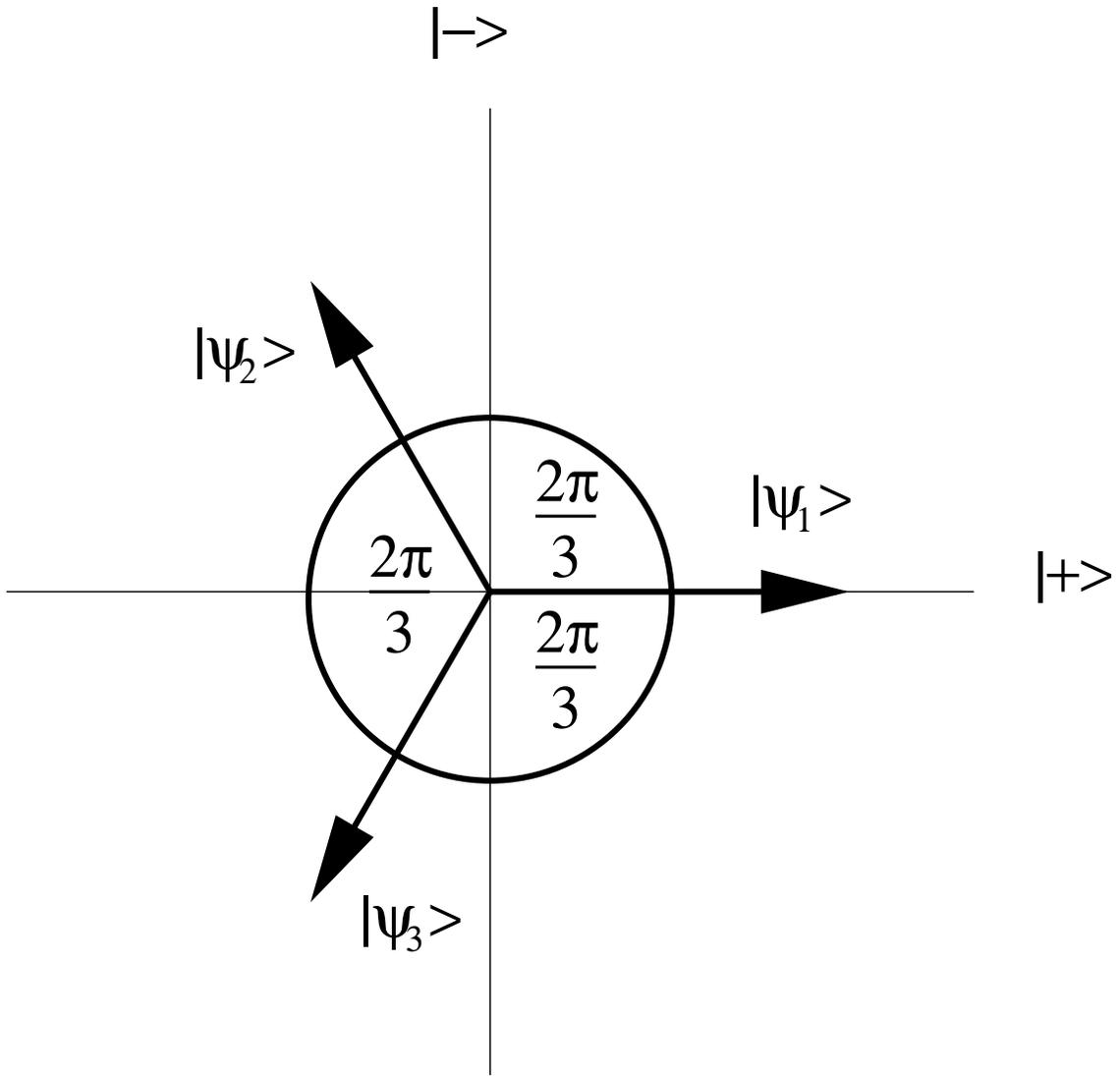}}
 \vspace{1cm}
 \caption{ Illustration of states forming the `trine'
ensemble. The states $|{\psi}_{j}{\rangle}$, where $j=1,2,3$, are
linear combinations of a pair of orthogonal states
$|{\pm}{\rangle}$. They have a real representation in this basis,
which enables us to represent them in a plane.  Each state is
obtained from its predecessor by a unitary transformation, this
being a counterclockwise rotation of $2{\pi}/3$ radians.  Three
consecutive such rotations are equivalent to the identity
operator.  As such, applying the rotation to the final state
$|{\psi}_{3}{\rangle}$ gives the initial state
$|{\psi}_{1}{\rangle}$.}
 \end{figure}

\newpage
\begin{figure}[tbp]
\epsfxsize15cm
 \centerline{\epsfbox{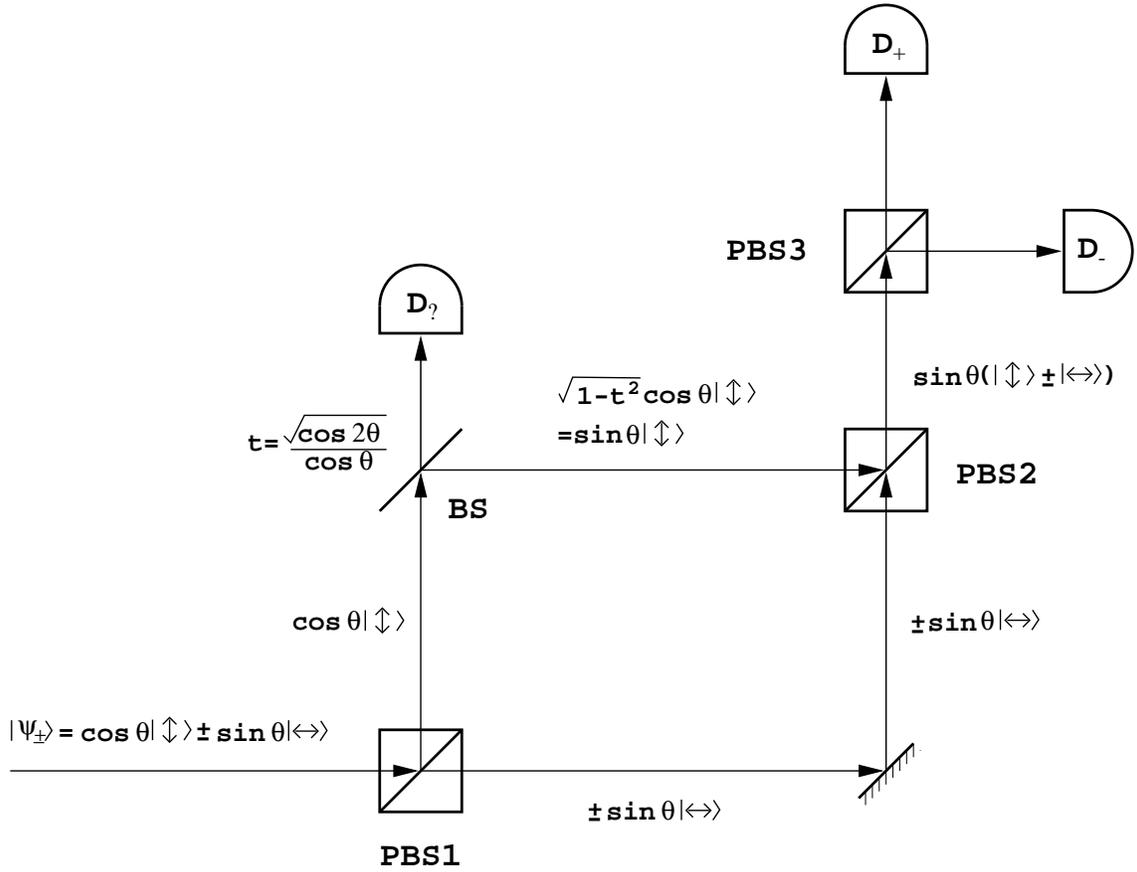}}
 \vspace{1cm}
 \caption{Optical measurement to unambiguously
discriminate between the non-orthogonal photon polarisation states
$|{\psi}_{\pm}{\rangle}={\cos}{\theta}|{\updownarrow}{\rangle}{\pm}{\sin}{\theta}|{\leftrightarrow}{\rangle}$
at the Ivanovic-Dieks-Peres bound.  The photon enters the
apparatus at the bottom left, where a polarising beamsplitter PBS1
transmits the horizontal component and reflects the vertical one.
The horizontal component propagates counter-clockwise to a second
polarising beamsplitter PBS2.  Meanwhile, the vertical component
encounters an ordinary beamsplitter BS with transmission
coefficient {\em t} given by Eq. (3.7).  If the photon is
transmitted, it will register at detector $D_{?}$.  Both states
$|{\psi}_{\pm}{\rangle}$ have equal probability of giving rise to
a detection here, so such outcomes give no information about the
state.  These are inconclusive results.  If, on the other hand,
this component is reflected, its amplitude is multiplied by the
reflection coefficient $r=\sqrt{1-t^{2}}$, which results in both
the vertical and horizontal components of both the states being
equal. When the horizontal and vertical components are recombined
by polarising beamsplitter PBS2, the resulting states,
corresponding to whether the initial state was
 $|{\psi}_{+}{\rangle}$ or $|{\psi}_{-}{\rangle}$.  These states
can thus be unambiguously discriminated by a final, suitably
oriented polarising beamsplitter PBS3.}
\end{figure}